\documentclass[final,5p,times,twocolumn,authoryear]{elsarticle}
\usepackage{mathtools}
\usepackage{algorithm}
\usepackage{tabularx} 
\usepackage{array}
\usepackage{longtable}
\usepackage{balance}
\usepackage{xspace}
\newcolumntype{C}[1]{>{\centering\let\newline\arraybackslash\hspace{0pt}}m{#1}}
\usepackage{graphicx} 
\usepackage{multirow}
\usepackage{booktabs}
\usepackage{float}
\usepackage{amsmath,amssymb,amsthm,amsfonts,xcolor, color,stmaryrd}
\usepackage{enumerate}
\usepackage[utf8]{inputenc}
\usepackage[english]{babel}
\usepackage{tikz}
\usetikzlibrary{matrix, shapes, arrows, positioning, chains}
\usepackage{ragged2e}
\usepackage{lipsum}
\usepackage{hyperref}
\usepackage{siunitx}
\usepackage{textgreek} 

\begin{document}
	\begin{frontmatter}
		
\title{Hybrid Expansion Cosmology in $f(T)$ Gravity: Late-Time Evolution and Observational Bounds}

\author[inst1]{Rajalaxmi Jena\corref{cor1}}
\ead{rajalaxmijena066@gmail.com} 
\author[inst2]{Vishal M C}
\ead{vishal.mc2023@vitstudent.ac.in}
\author[inst1]{Sankarsan Tarai}
\ead{sankarsan.tarai@vit.ac.in}

\address[inst1]{Department of Mathematics, School of Advanced Sciences,
	Vellore Institute of Technology, Chennai-600127, India}

\address[inst2]{Department of Physics, School of Advanced Sciences,
	Vellore Institute of Technology, Chennai-600127, India}

\begin{abstract}
The cosmological expansion history of the exponential $f(T)$ gravity model proposed by Linder is determined from a modified Friedmann equation that generally does not admit a closed-form analytical solution for the Hubble parameter, requiring repeated numerical evaluations in cosmological analyses. Motivated by this limitation, we present an analytical reconstruction of the model by introducing a Lambert--W-based hybrid expansion ansatz for the Hubble parameter. The free parameters of the reconstruction are first determined through a least-squares optimization by minimizing the deviation from the numerical solution of the modified Friedmann equation, yielding a maximum relative difference of approximately $1.08\%$ over the redshift range $0\leq z\leq4$. The reconstructed model is subsequently treated as an independent analytical cosmological model and constrained using recent late-time observations, including Cosmic Chronometer (CC), DESI DR2 Baryon Acoustic Oscillation (BAO), Pantheon+SH0ES, and Union3 Type Ia supernova datasets through Markov Chain Monte Carlo (MCMC) analyses. The resulting parameter constraints are mutually consistent across the individual datasets and become significantly tighter for the combined likelihood. Furthermore, various cosmological diagnostics demonstrate that the reconstructed model successfully describes the transition from an early decelerated universe to the present accelerated epoch while remaining in close agreement with the standard $\Lambda$CDM cosmology at late times. These results demonstrate that the proposed hybrid ansatz accurately reproduces the numerical exponential $f(T)$ expansion history while providing a computationally efficient analytical framework for cosmological parameter estimation and future observational studies.
\end{abstract}


\end{frontmatter}

\section{Introduction}
Modern cosmology continues to confront profound challenges in providing a coherent explanation for the accelerated expansion of the late-time universe. Cosmological observations from various supernova-like type $\mathrm{I}a$ Supernovae (SNe)$\mathrm{I}a$ \cite{Riess1998, Riess2004}, the Cosmic Microwave Background Radiation(CMBR) \cite{E.Komatsu2009}, and Large Scale Structure (LSS) \cite{Eisenstein2005, Tegmark2004, Seljak2005} collectively indicate that the universe is presently experiencing an accelerated expansion during its late-time evolution \cite{Cai2016, Riess1998, Hinshaw2013, Perlmutter1999, Wang2008}. This phenomenon is generally attributed to an elusive component known as dark energy, which is characterized by a fluid with negative pressure. It is widely accepted that dark energy constitutes nearly two-thirds of the total energy of the universe. Recent observational evidence indicates that the combined contribution of dark energy and dark matter constitutes roughly $95-96\%$ of the total energy budget of the universe \cite{Weinberg1989}. Among the numerous theoretical proposals put forward to describe dark energy, the cosmological constant ($\Lambda$) remains the most straightforward and widely supported candidate. The resulting $\Lambda\text{CDM}$ model demonstrates remarkable agreement with a wide range of observational data. Although the cosmological constant provides a simple explanation for cosmic acceleration, it is accompanied by two significant theoretical issues, commonly referred to as the fine-tuning problem and the coincidence problem. \cite{DiValentino2021, Edmund2006, Padmanabhan2003, Carroll2001}. These persistent issues have encouraged the investigation of alternative directions in cosmology, prompting the development of models that extend beyond the conventional framework.\\
\\
To address these challenges, researchers have explored a variety of possible directions. One line of work introduces additional or non-standard forms of matter and energy to account for the observed acceleration. Another, more structural approach involves modifying Einstein’s general relativity itself. The latter has drawn considerable attention, mainly because it offers the possibility that the accelerated expansion could arise from the gravitational sector, without the need to introduce a separate dark energy component. General relativity, despite its remarkable success in explaining gravitational phenomena over a wide range of scales, does face difficulties when applied to late-time cosmic acceleration. This has motivated the development of modified gravity theories that attempt to extend or go beyond the standard framework \cite{Clifton_2012, Nojiri_2017}. In many cases, these theories are constructed by altering the curvature-based Einstein–Hilbert action, leading to modified field equations that can naturally accommodate an accelerating universe. Prominent examples include $f(R)$ gravity \cite{Sotiriou2010, DeFelice2010, Nojiri2011}, $f(R, T)$ gravity \cite{Harko2011, Myrzakulov2012}, $f(R,L_{m})$ gravity\cite{Harko_2010, KalpanaDevi2024,KalpanaDevi2025} $f(G)$ gravity \cite{De.Felice2009}, $f(R, G)$ gravity \cite{Nojiri_2005}, $f(Q)$ gravity \cite{Koussour2022, Heisenberg2024}, $f(T,B)$ gravity \cite{Bahamonde_2015,Bahamonde2023}, $f(Q, T)$ gravity  \cite{XuYixin2019} and $ f(T)$ gravity \cite{Bahamonde2023, Bengochea2009, Aldrovandi2012, Cardone2012,PhysRevD.85.104036,PhysRevD.88.084042}. Each of these approaches provides a distinct perspective on gravitational interactions and the overall evolution of the universe. The explanations put forward to account for the accelerated expansion of the universe can be divided into two principal categories. One possibility is to extend the matter sector by introducing additional dynamical fields, such as scalar fields (both canonical and phantom) or even vector fields, which can drive the acceleration \cite{Copeland_2006, Bassett2006}. The other approach is to reconsider the gravitational framework itself. In this case, gravity can be formulated in different but equivalent geometric ways, based on curvature, torsion, or non-metricity \cite{Capozziello2011, DiValentino2025}. Within this latter category, gravity can be described through three equivalent geometrical frameworks, such as curvature, torsion, and non-metricity \cite{Beltran2019}.\newline

Among these, torsion-based modifications offer a particularly promising direction due to their distinct geometric structure and the novel physical insights they provide. In particular, the torsion-based formulation has been extensively studied in recent years. In contrast to curvature-based gravity, where gravitational effects arise from the curvature of spacetime, the teleparallel formulation attributes gravitation to spacetime torsion. In this framework, gravity is described through the Teleparallel Equivalent of General Relativity (TEGR), which provides a formulation dynamically equivalent to General Relativity at the level of the field equations, while employing torsion instead of curvature as the fundamental geometrical quantity. This viewpoint opens new avenues for constructing modified theories of gravity that are dynamically distinct from their curvature-based counterparts. In particular, the extension to $f(T)$ gravity has emerged as one of the most extensively studied modifications, owing to its ability to modify gravitational dynamics while preserving second-order field equations\cite{Bahamonde:2021gfp}. This feature makes it mathematically simpler and computationally more tractable compared to curvature-based extensions such as $f(R)$ gravity. Moreover, $f(T)$ gravity provides a compelling framework for explaining the observed late-time acceleration of the universe without introducing an explicit dark energy component. Recent studies have investigated its cosmological implications and observational consistency. Several recent studies have explored the cosmological implications of extended teleparallel gravity models \cite{Lohakare2023, Kofinas_2014}. In particular, Lohakare {\it et al.} \cite{Lohakare2023} investigated an extended $f(T, T_G)$ gravity framework constrained by observational datasets, demonstrating that the model successfully reproduces the cosmic expansion history. Their analysis predicts a smooth transition from an early decelerating phase to a late-time accelerating universe with a quintessence-like behavior, while also satisfying the energy conditions at the background level. SK Tripathy {\it et al.} \cite{Tripathy2024} develop a scalar-field extension of teleparallel gravity to model bounce cosmology and test its physical consistency. In \cite{Suresh_2023}, the authors investigate the viability of the power-law   $(f(T))$ gravity model using a combination of recent cosmological observations, including the Planck 2018 Cosmic Microwave Background (CMB), Pantheon Type Ia Supernovae, Baryon Acoustic Oscillations (BAO), and Big Bang Nucleosynthesis (BBN) datasets. To perform a comprehensive analysis, they consider both the background expansion history of the universe and the evolution of linear cosmological perturbations. This approach enables a more stringent comparison between the theoretical predictions of the model and current observational data, leading to tighter constraints on the model parameters\cite{Nunes_2018}. The study explores the impact of modified $f(T)$ gravity on the evolution of cosmic structures and the CMB through scalar perturbations. The analysis shows that the model agrees well with current cosmological observations while offering a potential resolution to the Hubble tension.
Within the torsion-based formulation, the Teleparallel Equivalent of General Relativity (TEGR) provides an alternative description of gravitation. In this framework, torsion replaces the curvature of the fundamental quantity governing gravitational interactions, and the tetrad fields serve as the fundamental dynamical variables instead of the metric tensor. The tetrad basis ${e_A(x^\mu)}$, where A = 0,1,2,3 labels the tangent-space coordinates and $\mu$ = 0,1,2,3 denotes spacetime indices, forms a set of orthonormal vectors. Each Vector $e_A$ can be decomposed into a coordinate basis and produces the components ${e_A^\mu}\partial_\mu$. These tetrad fields form an orthonormal set of four linearly independent basis vectors defined in the tangent space at each point in spacetime. Hence, the orthogonality condition becomes
\begin{equation}\label{Eq1} 
g_{\mu\nu} = \eta_{AB} \, e^A_{\ \mu} \, e^B_{\ \nu}.
\end{equation} \newline
Where $g_{\mu\nu}$ is the metric tensor and $\eta_{AB}=\mathrm{diag}(1,-1,-1,-1)$ is the Minkowski metric. The co-frame $\{e^\mu _A\}$ represents the inverse of the basis, enabling us to obtain the relation as\[e^\mu_{A} e^A_{\nu} = \delta^\mu_\nu\]  and \[e^\mu_{\ A}e^B_{\ \mu} = \delta^B_A\]

An important extension of TEGR is provided as $f(T)$ gravity, which has been proposed as a viable framework for explaining late-time cosmological phenomena in the universe. In contrast to GR, where gravitational interactions are governed by the curvature scalar $R$, the teleparallel formulation is based on the torsion scalar $T$. Within this approach, the $f(T)$ gravity is constructed by promoting the torsion scalar to a generalized function, thereby modifying the standard teleparallel Lagrangian density. As a result, the theory extends the TEGR action from a linear dependence on $T$ to a more general form involving $T + f(T)$, allowing for a richer corresponding action can be written as:
\begin{equation}
S = \frac{1}{16 \pi G} \int d^4x \, e \left[T+f(T) + \mathcal{L}_m\right]
\label{eq: action}
\end{equation}

In the above action, $\mathcal{L}_m$ the total matter Lagrangian, and $e = \det[e^A_{\ \mu}] = \sqrt{-g}$ represent the determinant of the tetrad fields. Throughout this work, we adopt natural units such that $8\pi G= \kappa^2=1$. Within the teleparallel framework, the gravitational action is constructed from the torsion scalar $T$, in contradiction to GR, which is formulated using the torsionless Levi-Civita connection and the curvature scalar $R$. Teleparallel Gravity, on the other hand, employs the curvature-free Witz-connection to describe gravitational interaction. As a result, gravity is attributed to space-time torsion rather than curvature. In this formulation, the fundamental connection is defined as
\begin{equation}
    {\hat \Gamma}^\lambda_{\ \mu\nu} \equiv e^\lambda_{\ A} \, \partial_\mu e^A_{\ \nu}
\end{equation}
Within the Weitzenböck connection, the curvature tensor identically vanishes, implying that spacetime is described by zero curvature while torsion remains non-zero. Based on this framework, the torsion tensor can be expressed as follows
\begin{equation}
T^\lambda_{\ \mu\nu} \equiv \hat{\Gamma}^\lambda_{\ \nu\mu} - \hat{\Gamma}^\lambda_{\ \mu\nu} = e^\lambda_{\ A} \left( \partial_\mu e^A_{\ \nu} - \partial_\nu e^A_{\ \mu} \right).
\label{eq:torsion}
\end{equation}
Where ${\hat \Gamma}^\lambda_{\ \mu\nu}  $ is the connection coefficient of the Witzenberg connection.\\
Further, the contortion tensor can be defined as 
\begin{equation}
 K^{\mu\nu}_{\ \ \ \rho} \equiv \frac{1}{2} \left( T^{\nu\mu}_{\ \ \ \rho} + T_\rho^{\ \mu\nu} - T^{\mu\nu}_{\ \ \ \rho} \right)
\end{equation}
From the above, we can define the superpotential tensor 
\begin{equation}
S_{\rho}^{\ \ \mu\nu} \equiv \frac{1}{2} \left( K^{\mu\nu}_{\ \ \ \rho} + \delta^\mu_\rho T^{\alpha \nu}_{\ \ \alpha} - \delta^\nu_\rho T^{\alpha\mu}_{\ \ \alpha} \right),
\label{eq: superpotential tensor}
\end{equation}
Using the above, we can write the teleparallel Lagrangian density, which is the torsion scalar.The torsion scalar is defined through an appropriate contraction of the torsion tensor. We can write it as 
\begin{equation}
T \equiv S_{\rho}^{\ \ \mu \nu} T^{\rho}_{\ \ \mu\nu}
\label{eq:torsionscalar}
\end{equation}

 A considerable body of work has been devoted to exploring cosmological solutions within the framework of  $ f(T)$ gravity, particularly in relation to the dynamical evolution of the universe \cite{Paliathanasis2016}. In addition, investigations into the thermodynamic properties of these models have been carried out \cite{Salako2013}, while classical energy conditions have been examined to assess their physical viability. A key feature of $f(T)$ gravity is that the resulting field equations are of second order, unlike those in $f(R)$ gravity, which generally involve fourth-order differential equations. This feature makes the theory comparatively simpler and more tractable for cosmological applications.
 The $f(T)$ framework has been widely employed in both astrophysical and cosmological studies, offering an alternative route to explain the late-time accelerated expansion of the universe without explicitly introducing a dark energy component. The gravitational field equations are derived by performing a variation of the action \eqref{eq: action} with respect to the vierbein, leading to 
\begin{equation}
\begin{split}
e^{-1} \partial_\mu \left(e\, e^\rho_{\ A} S_\rho^{\ \mu\nu} \right) [1 + f_T] 
+ e^\rho_{\ A} S_\rho^{\ \mu\nu} \, \partial_\mu(T)\, f_{TT}
\\- e^\lambda_{\ A} T^\rho_{\ \mu\lambda} S_\rho^{\ \nu\mu} [1 + f_T]
+ \frac{1}{4} e^\nu_{\ A} [T + f(T)] = 4\pi G\, e^\rho_{\ A} \mathcal{T}_\rho^{\ \nu},
\end{split}
\label{eq:fteq}
\end{equation}

We consider $f$ to be a function of the torsion scalar, written as $f(T)$, where the quantities $f_T=\frac{df}{dT}$ and $f_{TT}=\frac{d^2 f}{dT^2}$ represent its first and second derivatives with respect to $T$, respectively. In addition, $\mathcal{T}^\rho_{\ \nu} $ represents the total matter energy-momentum tensor constructed by the Lagrangian matter field.\\

The structure of this paper is outlined as follows. In
Section 2, we introduce the hybrid scale factor ansatz
within the framework of teleparallel $f(T)$ gravity and
derive the corresponding Hubble parameter in terms of
the Lambert-$W$ function. Section 3 presents the
exponential $f(T)$ gravity model proposed by Linder,
along with the power-law and exponential $f(T)$
parameterizations considered for comparison. In
Section 4, we validate the proposed analytical
Lambert-$W$ hybrid reconstruction by comparing its
expansion history with the numerical solution of the
original exponential $f(T)$ model and with other
analytical parameterizations. Section 5 is devoted to
the observational constraints on the reconstructed
model using Cosmic Chronometer (CC), DESI DR2
Baryon Acoustic Oscillation (BAO), Pantheon+SH0ES,
and Union3 Type Ia supernova datasets, together with
their combined analysis. Section 6 investigates the
physical and dynamical properties of the reconstructed
model through the deceleration parameter, effective
equation-of-state parameter, statefinder diagnostics,
$Om(z)$ diagnostic, and the energy conditions. Finally,
Section 7 presents the conclusion and discussion,
summarizing the main results and outlining possible
directions for future investigations.

\section{ Hybrid Scale Factor Ansatz} 
For the analysis of the cosmological framework, we adopt the spatially flat Friedmann-Lemaître-Robertson-Walker (FLRW) metric expressed in Cartesian coordinates.
\begin{equation}
ds^2 = dt^2-a(t)^2 \left( dx^2 + dy^2 + dz^2 \right),
\label{eq:flrw_cartesian}
\end{equation}

Here, \( a(t) \) is the scaling factor, and the tetrad is chosen as 
\begin{equation}
   e^A_{\ \mu}=diag(1,a(t),a(t),a(t))  .
\end{equation}
For the spatially flat FLRW geometry, the torsion scalar is given by
\begin{equation}
T=-6H^2.
\end{equation}
Varying the action with respect to the tetrad yields the modified Friedmann equations

\begin{equation}
3H^2 = \kappa\rho + \frac{1}{2} \left(2T f_T - f \right)
\end{equation}

\begin{equation}
2\dot{H} = -\frac{\kappa({\rho+p})}{1+f_T+2Tf_{TT}}
\end{equation}
The Hubble parameter is defined as $H=\frac{\dot{a}}{a}$, which characterizes the expansion rate of the universe, where the overdot denotes differentiation with respect to cosmic time $t$. Here, $\rho$ denotes the energy density and $p$ corresponds to the pressure. The total energy-momentum tensor includes contributions from both matter and radiation components. 

To investigate the cosmological behaviour of the proposed exponential $f(T)$ model, we rewrite the modified Friedmann equation in terms of the redshift $z$, defined by $1+z=a_0/a$, where the present value of the scale factor is normalized to $a_0=1$. We introduce the dimensionless Hubble parameter as
\begin{equation}
E(z)=\frac{H(z)}{H_0},
\end{equation}
where $H_0$ is the present-day Hubble parameter. Assuming a pressureless matter component, the matter density satisfies the standard conservation relation
\begin{equation}
\rho_m=\rho_{m0}(1+z)^3.
\end{equation}

The cosmological evolution considered in the present work is based on the hybrid scale factor (HSF) ansatz, originally introduced \cite{aydiner2022late,akarsu2014cosmology,Mishra:2015jja,Mishra:2017zlk}, which combines a power-law expansion with an exponential expansion. The scale factor is assumed to take the form \cite{Behera:2026yya}
\begin{equation}
a(t)=t^{\beta}e^{\lambda t},
\label{eq:hsf}
\end{equation}
where $\beta>0$ and $\lambda>0$ are constant model parameters. The power-law term dominates the early-time evolution and describes the decelerating expansion associated with the matter-dominated era, whereas the exponential term governs the late-time evolution, naturally leading to an accelerated expansion. Consequently, the hybrid scale factor provides a smooth transition from a decelerating universe to the present accelerating phase using only two free parameters.

Differentiating Eq.~(\ref{eq:hsf}) with respect to cosmic time, the corresponding Hubble parameter is obtained as
\begin{equation}
H(t)=\frac{\dot{a}}{a}=\frac{\beta}{t}+\lambda.
\label{eq:Ht}
\end{equation}
 Using the relation between the scale factor and redshift, $a(t)=\frac{1}{1+z}$, this expression can be reformulated in terms of the redshift as follows  
\begin{equation}
H(z) = \lambda\left[ 1+\frac{1}{W[{\frac{\lambda}{\beta}(1+z)^{\frac{-1}{\beta}}}]}\right]
\label{Eq:hubble}
\end{equation}
where 
\begin{equation}
{\frac{\lambda}{\beta}(1+z)^{\frac{-1}{\beta}}}=W\left[\frac{\lambda}{\beta}(1+z)^{\frac{-1}{\beta}}\right]e^{W\left[\frac{\lambda}{\beta}(1+z)^{\frac{-1}{\beta}}\right]}
\end{equation}
Another important geometrical quantity is the deceleration parameter
\begin{equation}
q(t) = -1 + \frac{\beta}{(\beta+\lambda t)^2}.
\end{equation} 
The deceleration parameter can be expressed in a parametrized form as
\begin{equation}
q(z) = -1+ \frac{1}{\beta[1+W[\frac{\lambda}{\beta}(1+ z)^{\frac{-1}{\beta}}]]^2}
\end{equation}

The Statefinder parameter $r$, which is equivalent to the cosmological jerk parameter, is given by
\begin{equation}
r(t)=1-\frac{3\beta}{(\beta+\lambda t)^2}
+\frac{2\beta}{(\beta+\lambda t)^3}.
\end{equation}

Expressed as a function of redshift, the Statefinder parameter becomes
\begin{equation}
r(z)=1-\frac{3}{\beta[{1+W[\frac{\lambda}{\beta}(1+z)^{\frac{-1}{\beta}}]}^2}+\frac{2}{\beta^2[{1+W[\frac{\lambda}{\beta}(1+z)^{\frac{-1}{\beta}}]}^3}
\end{equation}

The parametrized form of $H(z)$ in Eq.~(\ref{Eq:hubble}) describes the redshift evolution of the expansion rate. This raw function does not inherently satisfy the standard normalization condition at the present epoch, $H(z = 0) = H_0$. Given that $H_{0}$ is tightly constrained by both the Planck mission and local distance ladder measurements \cite{Planck:2018vyg} and \cite{Riess2022}, any physically consistent model must reproduce this value at $z = 0$. To ensure this consistency, we implement a normalization procedure that rescales the model’s amplitude while preserving its intrinsic redshift dependence:
\begin{equation}
H_{\text{norm}}(z) = H_0 \frac{H_{\text{mode}l}(z)}{H_{\text{model}}(0)},
\end{equation}

This transformation ensures that $H_{\text{norm}}(0) = H_0,$ aligning the model with current observational benchmarks without altering its fundamental dynamics. Now the normalized equation has become,
\begin{equation}
E(z)
=
\frac{H(z)}{H_0}
=
\frac{
\lambda
\left[
1+\dfrac{1}{W\!\left(\dfrac{\lambda}{\beta}(1+z)^{-1/\beta}\right)}
\right]
}{
\lambda
\left[
1+\dfrac{1}{W\!\left(\dfrac{\lambda}{\beta}\right)}
\right]
},
\label{eq:normalized_H1}
\end{equation}\\
This then reduces to,\\
\begin{equation}
E(z)
=
\frac{
1+\dfrac{1}{W\!\left(\dfrac{\lambda}{\beta}(1+z)^{-1/\beta}\right)}
}{
1+\dfrac{1}{W\!\left(\dfrac{\lambda}{\beta}\right)}
},
\label{eq:normalized_H}
\end{equation}\\
where $E(z)=H(z)/H_0$ is the dimensionless Hubble parameter, $W(x)$ denotes the principal branch of the Lambert-$W$ function, and the normalization ensures that $E(0)=1$, thereby satisfying the present-day boundary condition $H(0)=H_0$. Equation~(\ref{eq:normalized_H}) provides a closed-form analytical expression for the normalized Hubble parameter.\\

\section{Cosmological $f(T)$ Models}

Among the various modified teleparallel gravity models, exponential $f(T)$ gravity has attracted considerable attention for introducing smooth deviations from TEGR while remaining compatible with cosmological observations and local gravity constraints \cite{Linder2010, Cai2016}, and has been successfully tested against Type Ia supernovae, baryon acoustic oscillations, cosmic chronometers, and CMB data \cite{Basilakos2013, Nunes2016, Nunes_2018, PhysRevD.107.063529}. Motivated by these features, we consider the exponential $f(T)$ model proposed by Linder, but instead of its conventional cosmological evolution, we reconstruct the background dynamics using a hybrid scale factor that naturally describes the transition from an early decelerating universe to the present accelerating phase \cite{Akarsu2014, Akarsu2019}. This allows us to investigate whether the exponential $f(T)$ model preserves its cosmological viability under an alternative expansion history, and to compare the resulting parameter constraints with those reported in previous studies. In this work, we consider the exponential $f(T)$ gravity model proposed by Linder \cite{Linder2010}, given by
\begin{equation}
	f(T) = \alpha T_0\left[1-e^{-b\sqrt{\frac{T}{T_0}}}\right]
\end{equation}
where $b$ and $\alpha$ are model parameters. The present-day value of the torsion scalar is $T_0 =-6H_0^2$. From the above form of $f(T)$, we compute the first and second derivatives of \( f(T) \) with respect to \( T \). The first derivative is
  \[
  f_T =\frac{df}{dT} =\frac{b\alpha\sqrt{\frac{T}{T_0}}T_0(e^{-b\sqrt{\frac{T}{T_0}}})}{2T},
  \]
  The second derivative is
  \[
  f_{TT}=\frac{d^2f}{dT^2}=-\frac{b\alpha e^{-b\sqrt{\frac{T}{T_0}}}(bT+\sqrt{\frac{T}{T_0}}T_0)}{4T^2}
  \]

  By inserting the exponential $f(T)$ function, its derivative with respect to the torsion scalar, and the matter density evolution into the modified Friedmann equation, the corresponding dimensionless expression is obtained as

\begin{equation}
E^2(z)=\Omega_{m0}(1+z)^3-\alpha\left[-1+e^{-bE}(1+bE)\right].
\label{eq:linder_raw}
\end{equation}

where $\Omega_{m0}$ denotes the present-day matter density parameter, while
$\alpha$ and $b$ characterize the exponential correction introduced by the
adopted $f(T)$ gravity model. The first term on the right-hand side corresponds
to the standard contribution from pressureless matter, whereas the second term
represents the modification arising from the non-linear dependence of the
gravitational action on the torsion scalar.

The normalization constant $\alpha$ is determined by imposing the present-day
boundary condition $E(0)=1$. Evaluating Eq.~(\ref{eq:linder_raw}) at the
present epoch ($z=0$) gives

\begin{equation}
1=\Omega_{m0}
+\alpha\left[1-(1+b)e^{-b}\right].
\label{eq:linder_norm}
\end{equation}

Solving Eq.~(\ref{eq:linder_norm}) for $\alpha$ yields

\begin{equation}
\alpha=
\frac{1-\Omega_{m0}}
{1-(1+b)e^{-b}}.
\label{eq:alpha}
\end{equation}

Substituting Eq.~(\ref{eq:alpha}) into
Eq.~(\ref{eq:linder_raw}) gives the normalized
dimensionless Hubble equation

\begin{equation}
E^2(z)=\Omega_{m0}(1+z)^3+
(1-\Omega_{m0})
\left[
\frac{e^{b}-e^{-b(E-1)}(1+bE)}
{e^{b}-(1+b)}
\right],
\label{eq:dimensionless}
\end{equation}\\

Equation~(\ref{eq:dimensionless}) serves as the reference numerical expansion
history throughout the present work and is employed to validate the proposed
hybrid analytical reconstruction against the original exponential $f(T)$ model.\\

In addition to the exponential $f(T)$ model proposed by Linder, we consider two well-established functional forms that have been extensively investigated in the framework of modified teleparallel gravity. The first is the power-law model,
\begin{equation}
 f(T)=c(-T)^n 
\end{equation} 

This model has been widely studied owing to its simplicity and its ability to account for the late-time accelerated expansion of the universe without invoking an explicit dark energy component \cite{Bengochea2009, WuYu2010, Bengochea2010}. Here, $c$ and $n$ are constant model parameters. Following the substitution of the functional form into the modified Friedmann equation together with the matter conservation equation, the normalized Hubble equation becomes
\begin{equation}
E^2(z)=\Omega_{m0}(1+z)^3+
\left(1-\Omega_{m0}\right)E^{2n},
\label{eq:powerlaw}
\end{equation}
where $E(z)=H(z)/H_0$ and the second term encodes the deviation from standard teleparallel gravity through the power-law correction. The parameter $n$ determines the deviation from the Teleparallel Equivalent of General Relativity, with the standard cosmological evolution being recovered for the appropriate limiting value of the model parameter. Equation~(\ref{eq:powerlaw}) forms the basis for confronting the power-law $f(T)$ model with observational Hubble data and comparing its cosmological predictions with those of the exponential models considered in this work.\\

We also consider the exponential $f(T)$ model proposed by Bamba et al.\cite{Bamba_2011},
\begin{equation}
f(T)=\alpha T_0\left(1-e^{-pT/T_0}\right)    
\end{equation} This model has subsequently been constrained using various cosmological
observations \cite{Nesseris2013,Nunes2016,Cai2016}. Here, $\alpha$ and $p$ are constant model parameters, while
$T_0=-6H_0^2$ denotes the present-day torsion scalar. Substituting the above functional form and its corresponding derivative into the modified Friedmann equation, together with the matter conservation relation $\rho_m=\rho_{m0}(1+z)^3$, yields the dimensionless Hubble equation
\begin{equation}
E^2(z)=\Omega_{m0}(1+z)^3+
\alpha\left[1-\left(1+2pE^2\right)e^{-pE^2}\right].
\label{eq:linder}
\end{equation}
Here, $E(z)=H(z)/H_0$ is the normalized Hubble parameter. The first term represents the contribution from the pressureless matter component, whereas the second term accounts for the modification introduced by the exponential torsional correction. The parameters $\alpha$ and $p$ determine the amplitude and evolution of this correction, thereby influencing the expansion history of the universe. Equation (\ref{eq:linder}) is employed to investigate the cosmological evolution predicted by this model and to compare its cosmological evolution with the other $f(T)$ models considered in this work.\\

These models represent two distinct classes of modifications to teleparallel gravity and exhibit different cosmological dynamics. Therefore, comparing their predictions with those of the present model enables us to investigate the influence of different functional forms of $f(T)$ on the expansion history of the universe and to assess their consistency with the latest observational data.\\

\section{Validation of the Proposed Analytical Reconstruction}

The principal objective of the present work is to develop and validate an analytical reconstruction of the exponential $f(T)$ gravity model proposed by Linder\cite{Linder2010} through the introduction of a hybrid scale factor ansatz. 
Unlike the original exponential $f(T)$ model, whose cosmological expansion history is determined from a modified Friedmann equation that generally does not admit a closed-form analytical solution for $H(z)$, the proposed reconstruction provides an explicit analytical expression for the Hubble parameter, thereby eliminating the need for repeated numerical solutions during cosmological analyses during parameter estimation or cosmological diagnostics. This considerably reduces the computational complexity while preserving the physical behaviour of the original numerical model.\\

To validate the proposed analytical reconstruction, the Hubble parameter predicted by the hybrid ansatz is directly compared with the numerical solution of the exponential $f(T)$ model over the redshift interval $0\leq z\leq4$. The free parameters of the analytical reconstruction are determined through a least-squares optimization procedure in which the objective function minimizes the discrepancy between the analytical and numerical expansion histories. It should be emphasized that these optimized parameters are introduced solely for the purpose of validating the analytical reconstruction. They should therefore be distinguished from the cosmological parameters obtained later through Bayesian parameter estimation using observational datasets. Consequently, the present analysis evaluates the reconstruction capability of the proposed analytical ansatz rather than its observational performance.\\

\begin{figure}[H]
\centering
\includegraphics[width=0.8199\linewidth]{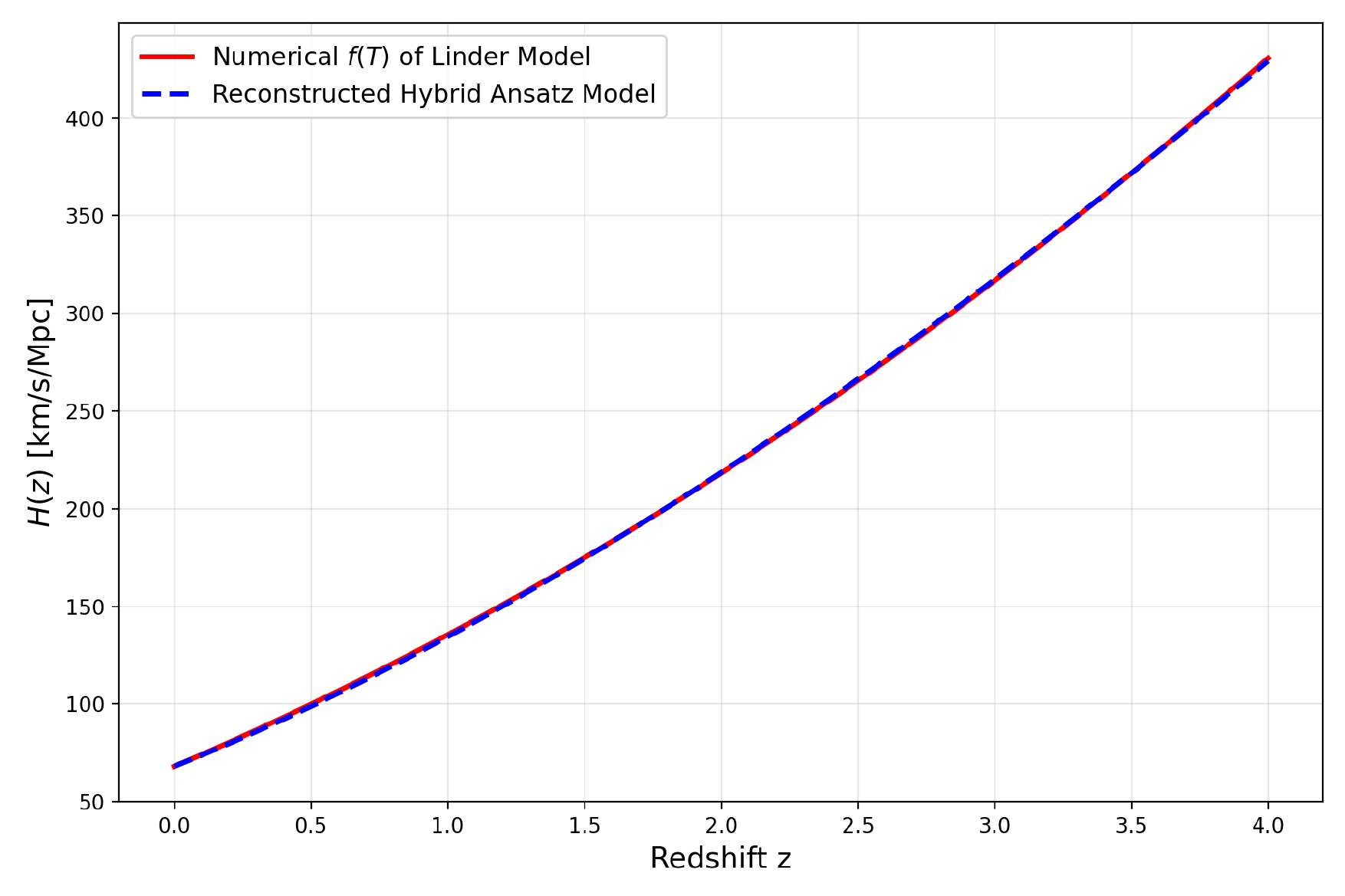}
\caption{Comparison between the numerical exponential $f(T)$ model and the proposed Lambert-W hybrid reconstruction.}
\label{fig:validation_Hz}
\end{figure}

Figure~\ref{fig:validation_Hz} illustrates the evolution of the Hubble parameter predicted by both the numerical exponential $f(T)$ model and the analytical reconstructed ansatz model. The two curves are almost indistinguishable over the entire redshift interval considered in this work, demonstrating that the proposed Lambert--W hybrid ansatz faithfully reproduces the cosmological expansion history of the original numerical solution. The excellent agreement extends from the present accelerated epoch to the matter-dominated era, indicating that the analytical reconstruction successfully captures the essential dynamical behaviour of the exponential $f(T)$ cosmology. Although the visual agreement already provides strong evidence for the validity of the reconstruction, a quantitative comparison is required to assess the accuracy of the approximation. The relative deviation between the analytical reconstruction and the numerical solution is therefore computed as

\begin{equation}
\delta(z)
=
\frac{H_{\rm Hybrid}(z)-H_{\rm Numerical}(z)}
{H_{\rm Numerical}(z)}
\times100\%.
\label{eq:relative_error}
\end{equation}

The corresponding relative error remains exceptionally small throughout the investigated redshift interval, with a maximum deviation of approximately $1.08\%$. Furthermore, the residuals remain uniformly distributed around zero without exhibiting any systematic trend, demonstrating that the analytical reconstruction neither consistently overestimates nor underestimates the numerical solution. Such a small reconstruction error confirms that the proposed hybrid ansatz provides an accurate analytical surrogate for the numerical exponential $f(T)$ model.\\

To further investigate the robustness of the proposed reconstruction, the resulting expansion history is compared with two widely used analytical parameterizations in teleparallel gravity, namely the power-law\cite{Bengochea2009} and exponential $f(T)$ model introduced by Bamba et al\cite{Bamba_2011}. This comparison enables an assessment of whether the proposed Lambert-W reconstruction provides an improved approximation to the numerical exponential $f(T)$ solution.\\

\begin{figure}[H]
\centering
\includegraphics[width=0.8199\linewidth]{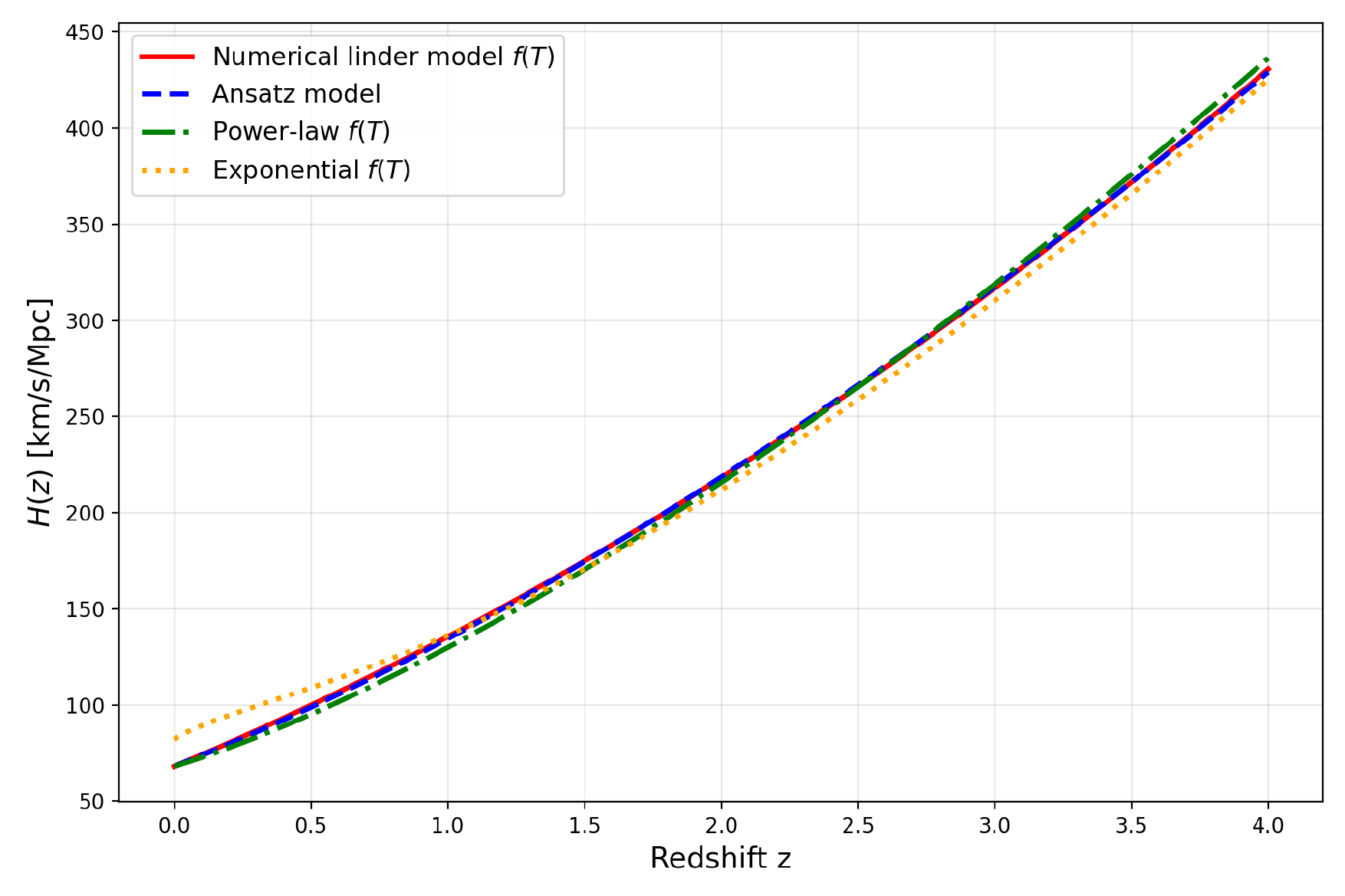}
\caption{Comparison of the numerical exponential $f(T)$ model with the proposed Lambert--W hybrid reconstruction, the power-law $f(T)$ model, and the exponential $f(T)$ parameterization.}
\label{fig:Models_comparison}
\end{figure}

Figure~\ref{fig:Models_comparison} compares the Hubble expansion histories predicted by the numerical exponential $f(T)$ model, the proposed Lambert--W reconstruction, and the two commonly adopted analytical parameterizations. While all three analytical models reproduce the overall expansion history, the proposed hybrid reconstruction remains almost indistinguishable from the numerical solution over the entire cosmological evolution. In contrast, the power-law parameterization exhibits noticeable deviations at intermediate redshifts, whereas the exponential parameterization departs more significantly from the numerical solution, particularly at lower redshifts.\\

The superiority of the proposed reconstruction is confirmed through this relative deviation analysis. The Lambert--W reconstruction exhibits a maximum relative difference of only approximately $1.08\%$, whereas the maximum deviations associated with the power-law and exponential parameterizations are approximately $4.78\%$ and $21.31\%$, respectively. These results demonstrate that the proposed hybrid ansatz provides the closest analytical representation of the numerical Linder exponential $f(T)$ expansion history. Furthermore, the larger deviations observed for the power-law and Bamba exponential models indicate that, although these models reproduce the general background evolution, they are unable to capture the numerical Linder cosmology with the same level of accuracy over the considered redshift range.\\

Table~\ref{tab:model_parameters} summarizes the cosmological and model parameters employed in the present analysis. The fiducial cosmological parameters $\Omega_{m0}\approx0.30$ and
$H_0\approx68~\mathrm{km\,s^{-1}\,Mpc^{-1}}$ are adopted as representative reference values consistent with the cosmological constraints reported by the Planck Collaboration \cite{Planck:2018vyg}. The exponential parameter of the numerical Linder $f(T)$ model is fixed to the representative value $b=1.0$. Since the primary objective of the present work is to validate the proposed hybrid analytical reconstruction rather than to constrain the parameters of the original exponential $f(T)$ model, fixing $b$ reduces the dimensionality of the reference model and enables a direct comparison between the analytical reconstruction and the corresponding numerical expansion history. The power-law parameter $n$, the Hybrid ansatz parameters $\lambda$ and $\beta$ and the Bamba exponential model parameters $p$ and $\alpha$ are determined through a least-squares optimization by minimizing the squared difference between the corresponding analytical Hubble functions and the numerical Hubble expansion of the Linder exponential $f(T)$ model over the redshift interval $0 \leq z \leq 4$. Consequently, these optimized parameters represent the optimal comparison values for reproducing the numerical Linder solution and should not be interpreted as observationally constrained cosmological parameters.

\begin{table}[ht]
\centering
\caption{Fiducial cosmological parameters and optimized model parameters adopted for the hybrid ansatz and comparison $f(T)$ gravity models used in the validation analysis.}
\label{tab:model_parameters}
\renewcommand{\arraystretch}{1.2}
\begin{tabular}{lc}
\hline
\textbf{Parameter} & \textbf{Value}\\
\hline
$\Omega_{m0}$ & $\approx0.30$  \\
$H_0$ & $\approx68.0$  \\
$b$ & 1.0  \\
$n$ & 0.4467  \\
$\lambda$ & 0.3232 \\
$\beta$ & 0.6720 \\
$p$ & 2.2515 \\
$\alpha$ & 1.6154  \\
\hline
\end{tabular}
\end{table}
It is important to note that the parameters obtained through the least-squares optimization are introduced exclusively for validating the analytical reconstruction. In the subsequent observational analysis, the Hybrid ansatz model parameters ($\lambda$ and $\beta$) are treated as free cosmological parameters and are independently constrained using CC, BAO, Pantheon+SH0ES, and Union3 observations through a MCMC analysis. Since the least-squares optimization minimizes the reconstruction error relative to the numerical solution, whereas the Bayesian analysis maximizes the observational likelihood, the resulting parameter values are not expected to be identical. These two optimization procedures address fundamentally different objectives and therefore provide complementary information regarding the performance of the proposed analytical model.\\

Overall, the validation presented in this section establishes that the proposed Lambert--W hybrid reconstruction accurately reproduces the numerical exponential $f(T)$ cosmology while simultaneously providing a substantial reduction in computational complexity. The excellent agreement between the analytical and numerical expansion histories, together with the maximum relative deviation of only $1.08\%$ and the superior performance compared with other analytical $f(T)$ parameterizations, demonstrates that the proposed reconstruction constitutes a reliable analytical surrogate for the numerical Linder model. These results provide a solid foundation for employing the reconstructed model in the subsequent observational parameter estimation and cosmological diagnostic analyses.

\section{Observational Constraints}

Following the successful validation of the proposed hybrid Lambert-W reconstruction against the numerical exponential $f(T)$ gravity model, we proceed to investigate its compatibility with current cosmological observations. While the previous section established the capability of the analytical ansatz to reproduce the numerical expansion history with high accuracy, the present analysis aims to determine whether the reconstructed model is also favoured by observational data. To this end, the free parameters of the hybrid model are treated as independent cosmological parameters and constrained using a joint analysis of the CC, BAO, Union3 and Pantheon+sh0es Type Ia Supernova datasets. The statistical analysis is carried out within the Bayesian framework using the affine-invariant MCMC sampler implemented in the \texttt{emcee} Python package\cite{ForemanMackey2013}. This methodology enables an efficient exploration of the multidimensional parameter space, yielding the posterior probability distributions, confidence regions, and best-fit estimates of the model parameters. The observational constraints obtained in this section provide an independent assessment of the cosmological viability of the proposed analytical reconstruction and form the basis for the subsequent cosmological diagnostics and model comparisons.\\

\subsection{Cosmic Chronometer Dataset}

The Cosmic Chronometer (CC) technique provides a direct and nearly model-independent determination of the Hubble expansion rate by measuring the differential age evolution of passively evolving galaxies. Unlike luminosity- or angular-distance probes, the CC method directly estimates the expansion history without assuming a particular cosmological model. The Hubble parameter is obtained from

\begin{equation}
H(z)
=
-\frac{1}{1+z}\frac{dz}{dt},
\label{eq:cc}
\end{equation}\\

where $dz/dt$ denotes the differential age evolution of galaxies observed at nearby redshifts. Owing to its direct dependence on galaxy age measurements, the Cosmic Chronometer technique provides one of the most reliable observational probes of the late-time expansion history of the universe.\\

In the present analysis, we employ a compilation of 31 Cosmic Chronometer measurements covering the redshift range $0.07 \leq z \leq 1.965$\cite{Lamaaoune:2025zhd}. The observed Hubble parameter measurements together with their associated uncertainties are used to construct the likelihood function for constraining the free parameters of the proposed hybrid cosmological model.\\

Assuming statistically independent Gaussian uncertainties, the corresponding chi-square statistic is given by

\begin{equation}
\chi_{\rm CC}^{2}
=
\sum_{i=1}^{31}
\left[
\frac{
H_{\rm obs}(z_i)
-
H_{\rm th}(z_i,\Theta)
}
{\sigma_i}
\right]^2,
\end{equation}

where $H_{\rm obs}(z_i)$ and $H_{\rm th}(z_i,\Theta)$ denote the observed and theoretical Hubble parameters at redshift $z_i$, respectively, $\sigma_i$ represents the observational uncertainty, and $\Theta=(H_0,\lambda,\beta)$ denotes the set of free model parameters.
The corresponding Gaussian likelihood is expressed as

\begin{equation}
\mathcal{L}_{\rm CC}
\propto
\exp
\left(
-\frac{\chi_{\rm CC}^{2}}{2}
\right),
\end{equation}

which is sampled using the affine-invariant MCMC ensemble sampler implemented in the \texttt{emcee} package to obtain the posterior distributions of the model parameters.\\

\begin{figure}[H]
    \centering
      \includegraphics[width=0.7199\linewidth]
      {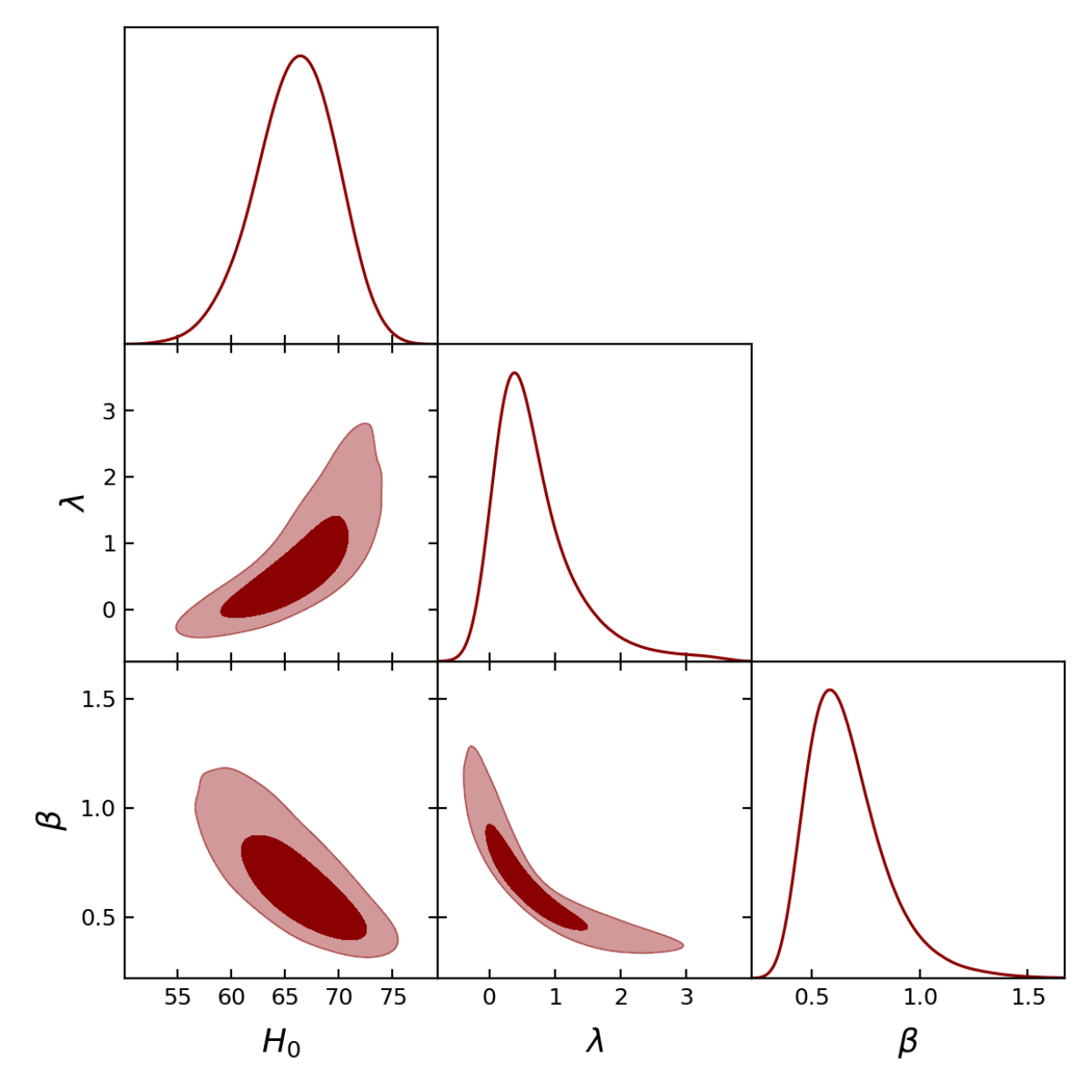}
    \caption{Marginalized posterior distributions and confidence contours for the free parameters $(H_0,\lambda,\beta)$ obtained from the Cosmic Chronometer (CC) dataset. The diagonal panels show the one-dimensional marginalized posterior distributions, whereas the off-diagonal panels present the corresponding two-dimensional posterior confidence contours. The shaded regions denote the 68\% and 95\% credible regions obtained from the MCMC analysis.}
    \label{fig: Hz_CC}
\end{figure}

Figure~\ref{fig: Hz_CC} presents the marginalized posterior distributions together with the two-dimensional confidence contours of the free parameters $(H_0,\lambda,\beta)$ obtained from the Cosmic Chronometer dataset. The posterior distributions indicate that the model parameters are well constrained by the observational data. A noticeable anti-correlation is observed between $\lambda$ and $\beta$, whereas $H_0$ exhibits a positive correlation with $\lambda$. The corresponding median parameter values together with their $68\%$ confidence intervals are summarized in Table~\ref{tab:mcmc_results}.

\subsection{Baryon Acoustic Oscillation Dataset}

\begin{table*}[t]
\centering
\footnotesize
\renewcommand{\arraystretch}{0.9}
\setlength{\tabcolsep}{4pt}

\caption{DESI DR2 BAO measurements adopted in this work. The official DESI DR2 GCcomb likelihood consists of 13 correlated BAO distance measurements together with the corresponding $13\times13$ covariance matrix. The observables are the transverse comoving distance ($D_M/r_d$), the Hubble distance ($D_H/r_d$), and the volume-averaged distance ($D_V/r_d$), where $r_d$ denotes the comoving sound horizon at the baryon drag epoch~\cite{DESI:2025zgx}.\\}
\label{tab:DESI_DR2_BAO}

\begin{tabular}{cccc}
\toprule
\textbf{Tracer} &
$\mathbf{z_{\rm eff}}$ &
\textbf{Observable} &
\textbf{Measured Value} \\
\midrule
BGS Bright        & 0.295 & $D_V/r_d$ & $7.9417 \pm 0.0761$ \\
LRG               & 0.510 & $D_M/r_d$ & $13.5876 \pm 0.1684$ \\
LRG               & 0.510 & $D_H/r_d$ & $21.8629 \pm 0.4289$ \\
LRG               & 0.706 & $D_M/r_d$ & $17.3507 \pm 0.1799$ \\
LRG               & 0.706 & $D_H/r_d$ & $19.4553 \pm 0.3339$ \\
ELG               & 0.934 & $D_M/r_d$ & $21.5756 \pm 0.1618$ \\
ELG               & 0.934 & $D_H/r_d$ & $17.6415 \pm 0.2010$ \\
QSO               & 1.321 & $D_M/r_d$ & $27.6009 \pm 0.3246$ \\
QSO               & 1.321 & $D_H/r_d$ & $14.1760 \pm 0.2246$ \\
QSO               & 1.484 & $D_M/r_d$ & $30.5119 \pm 0.7636$ \\
QSO               & 1.484 & $D_H/r_d$ & $12.8170 \pm 0.5180$ \\
Ly$\alpha$ Forest & 2.330 & $D_H/r_d$ & $8.6315 \pm 0.1011$ \\
Ly$\alpha$ Forest & 2.330 & $D_M/r_d$ & $38.9890 \pm 0.5317$ \\
\bottomrule
\end{tabular}
\end{table*}

Baryon Acoustic Oscillations (BAO) constitute one of the most reliable geometrical probes of the late-time expansion history of the universe. They originate from acoustic oscillations in the tightly coupled photon--baryon plasma before the epoch of recombination, leaving a characteristic comoving length scale that is preserved in the large-scale distribution of galaxies. Since the physical size of this acoustic scale is well determined, BAO observations provide a standard ruler for measuring cosmological distances over a wide range of redshifts.\\

In the present analysis, we employ the official DESI Data Release 2 (DR2) GCcomb BAO likelihood released by the DESI Collaboration~\cite{DESI:2025zgx}. The adopted compilation consists of 13 correlated BAO measurements spanning the effective redshift range
\[
0.295 \le z_{\rm eff} \le 2.330,
\]
including measurements of the transverse comoving distance ($D_M/r_d$), the Hubble distance ($D_H/r_d$), and the volume-averaged distance ($D_V/r_d$). The complete set of observables together with their associated uncertainties is summarized in Table~\ref{tab:DESI_DR2_BAO}. The statistical correlations among the measurements are fully incorporated through the covariance matrix provided by the DESI Collaboration.

For a spatially flat universe, the transverse comoving distance is given by

\begin{equation}
D_M(z)=c\int_0^z\frac{dz'}{H(z')},
\end{equation}

where $c$ denotes the speed of light. The corresponding Hubble distance is

\begin{equation}
D_H(z)=\frac{c}{H(z)},
\end{equation}

while the volume-averaged BAO distance is defined as

\begin{equation}
D_V(z)=\left[zD_M^2(z)D_H(z)\right]^{1/3}.
\end{equation}

The theoretical BAO observables are constructed using the dimensionless quantities

\[
\frac{D_M}{r_d}, \qquad
\frac{D_H}{r_d}, \qquad
\frac{D_V}{r_d},
\]

where the sound horizon at the baryon drag epoch is fixed throughout the analysis to

\[
r_d = 147.05~{\rm Mpc}.
\]

This value is adopted as a constant reference scale and is not varied during the parameter estimation\cite{Planck:2018vyg}. Consequently, the theoretical BAO predictions depend only on the cosmological expansion history described by the proposed hybrid model through the Hubble parameter $H(z)$.

The agreement between the theoretical predictions and the observational measurements is quantified using the Gaussian likelihood

\begin{equation}
\chi^2_{\rm BAO}
=
\Delta\mathbf{D}^{\,T}
\mathbf{C}^{-1}
\Delta\mathbf{D},
\end{equation}

where

\begin{equation}
\Delta\mathbf{D}
=
\mathbf{D}_{\rm th}
-
\mathbf{D}_{\rm obs},
\end{equation}

represents the residual vector between the theoretical and observed BAO quantities, and $\mathbf{C}$ denotes the full $13\times13$ covariance matrix supplied by the DESI Collaboration. The corresponding likelihood function is therefore

\begin{equation}
\mathcal{L}_{\rm BAO}
\propto
\exp\left(-\frac{\chi^2_{\rm BAO}}{2}\right).
\end{equation}

\begin{figure}[H]
    \centering
      \includegraphics[width=0.7199\linewidth]
      {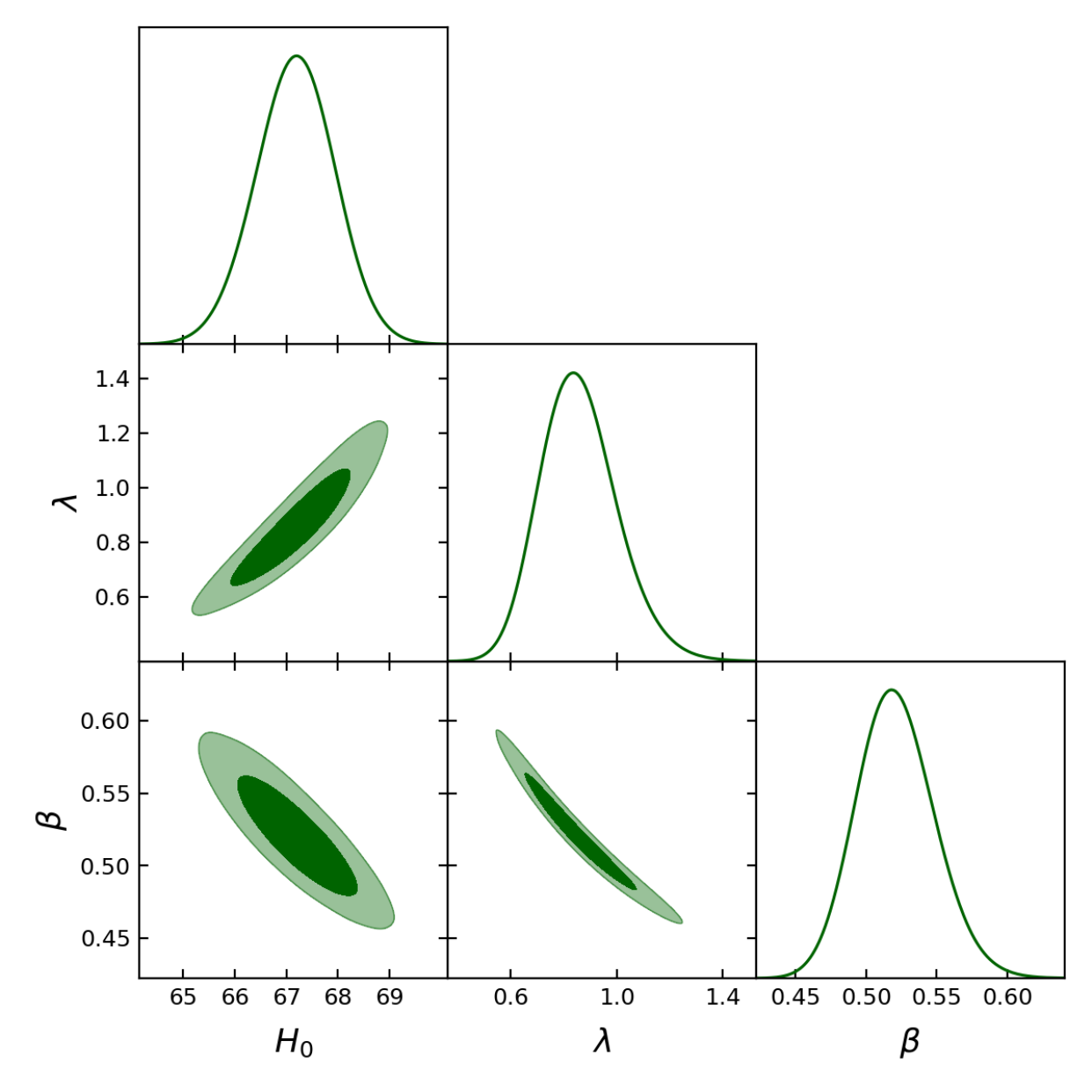}
    \caption{Marginalized posterior distributions and confidence contours for the free parameters $(H_0,\lambda,\beta)$ obtained from the Baryon Acoustic Oscillation (BAO) dataset. The diagonal panels show the one-dimensional marginalized posterior distributions, whereas the off-diagonal panels present the corresponding two-dimensional posterior confidence contours. The shaded regions denote the 68\% and 95\% credible regions obtained from the MCMC analysis.}
    \label{fig: BAO}
\end{figure}

Figure~\ref{fig: BAO} presents the best-fit predictions of the proposed hybrid model together with the DESI DR2 BAO measurements. The theoretical curves reproduce the observed BAO distance indicators over the entire redshift interval, while the shaded region represents the corresponding $1\sigma$ confidence interval obtained from the posterior distribution of the MCMC analysis. The close agreement between the reconstructed expansion history and the observational measurements demonstrates that the proposed hybrid model successfully reproduces the late-time geometrical constraints provided by the BAO standard ruler.

Finally, the BAO likelihood is combined with those of the Cosmic Chronometer (CC), Pantheon+SH0ES, and Union3 datasets to obtain joint constraints on the free parameters of the proposed hybrid cosmological model.

\subsection{Pantheon+SH0ES Dataset}

Type Ia Supernovae (SNe Ia) constitute one of the most precise observational probes for investigating the late-time expansion history of the universe owing to their remarkably uniform intrinsic luminosities. After appropriate empirical corrections, SNe Ia can be treated as standardizable candles, enabling accurate measurements of cosmological distances over a broad redshift range. In the present work, we employ the publicly available Pantheon+SH0ES compilation released by the Pantheon+ Collaboration~\cite{Brout:2022vxf, Riess2022}, which contains 1701 spectroscopically confirmed Type Ia supernovae spanning the redshift interval $0.001 \leq z \leq 2.26$. Among these, 77 supernovae are hosted in Cepheid-calibrated galaxies, thereby providing an absolute calibration of the supernova distance scale and allowing a direct constraint on the Hubble constant.\\

For a spatially flat Friedmann--Lemaître--Robertson--Walker (FLRW) universe, the comoving distance is given by

\begin{equation}
D_C(z)
=
c
\int_0^z
\frac{dz'}{H(z')},
\end{equation}

where $c$ denotes the speed of light and $H(z)$ is the Hubble parameter predicted by the cosmological model. The corresponding luminosity distance is

\begin{equation}
D_L(z)
=
(1+z)\,D_C(z),
\end{equation}

from which the theoretical distance modulus is obtained as

\begin{equation}
\mu_{\rm th}(z)
=
5\log_{10}
\left(
\frac{D_L(z)}{\rm Mpc}
\right)
+25.
\end{equation}

The luminosity distance is evaluated numerically by integrating the inverse Hubble parameter over a dense redshift grid using the cumulative trapezoidal rule, followed by cubic interpolation to obtain the theoretical prediction at the observed supernova redshifts.\\

Following the Pantheon+SH0ES analysis, the likelihood is constructed using the corrected apparent magnitudes together with the full statistical and systematic covariance matrix supplied by the collaboration. In addition to the cosmological parameters, the absolute magnitude of Type Ia supernovae, $M$, is treated as a nuisance parameter and is simultaneously constrained during the MCMC analysis. For Hubble-flow supernovae, the theoretical prediction is given by the model distance modulus, whereas for Cepheid-calibrated supernovae the corresponding host-galaxy Cepheid distance moduli are adopted.\\

The agreement between the theoretical model and the observational data is quantified through the Gaussian chi-square statistic,

\begin{equation}
\chi_{\rm SN}^{2}
=
\Delta\mathbf{m}^{\,T}
\mathbf{C}^{-1}
\Delta\mathbf{m},
\end{equation}

where

\begin{equation}
\Delta\mathbf{m}
=
\mathbf{m}_{\rm obs}
-
\mathbf{m}_{\rm model},
\end{equation}

is the residual vector between the observed and theoretical corrected apparent magnitudes, and $\mathbf{C}$ denotes the full $1701\times1701$ covariance matrix incorporating both statistical and systematic uncertainties. To improve the computational efficiency of the likelihood evaluation, the covariance matrix is factorized once using the Cholesky decomposition and subsequently employed throughout the MCMC analysis.\\

The corresponding likelihood function is therefore

\begin{equation}
\mathcal{L}_{\rm SN}
\propto
\exp
\left(
-\frac{\chi_{\rm SN}^{2}}{2}
\right).
\end{equation}

\begin{figure}[H]
    \centering
      \includegraphics[width=0.8199\linewidth]
      {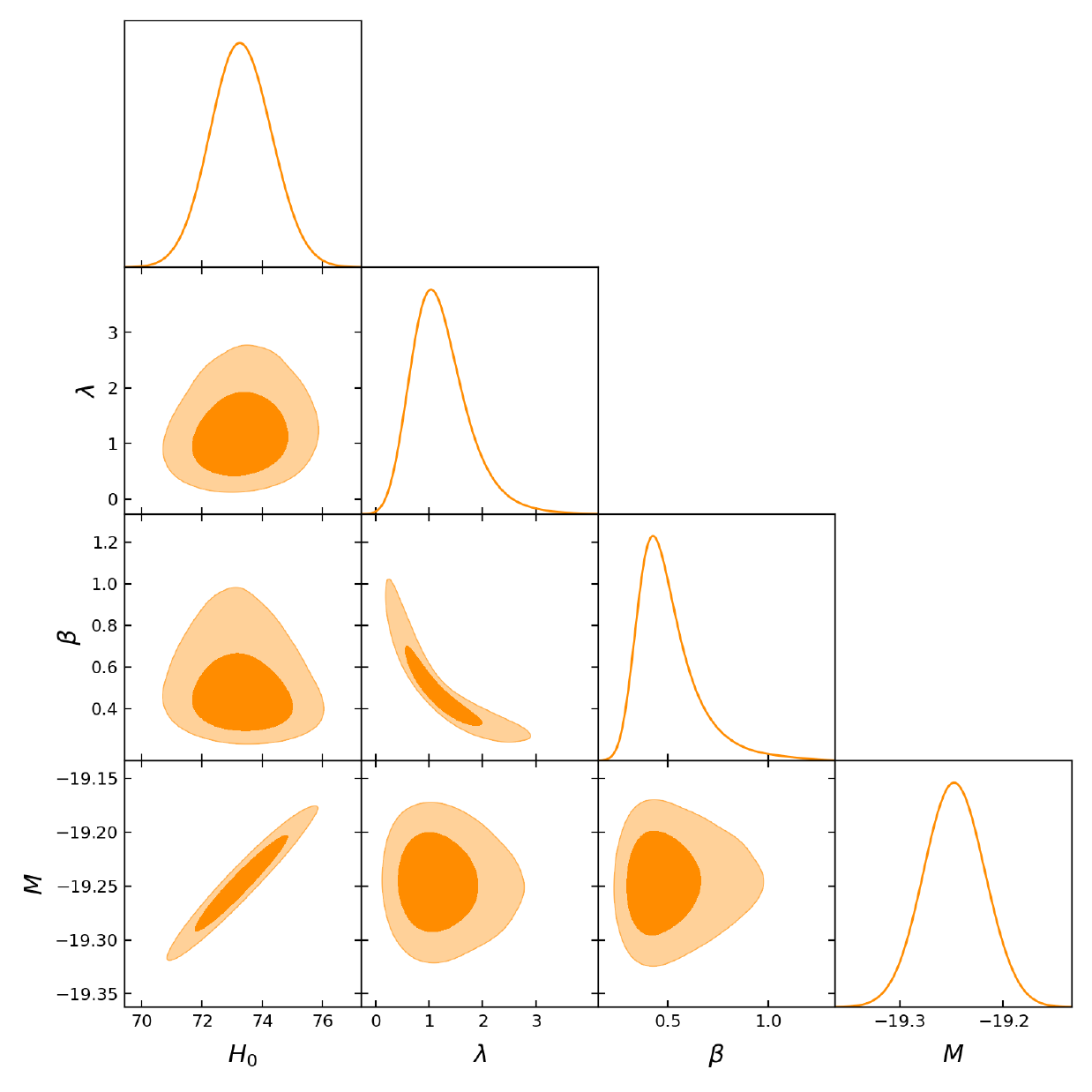}
    \caption{Marginalized posterior distributions and two-dimensional posterior contours for the free parameters $(H_0,\lambda,\beta,M)$ obtained from the Pantheon+SH0ES dataset. The diagonal panels show the one-dimensional marginalized posterior distributions, whereas the off-diagonal panels display the corresponding two-dimensional posterior contours. The shaded regions represent the 68\% and 95\% Bayesian credible regions obtained from the MCMC analysis.}
   \label{fig: Pantheon}
\end{figure}

Figure~\ref{fig: Pantheon} presents the marginalized posterior distributions together with the two-dimensional confidence contours of the free parameters $(H_0,\lambda,\beta,M)$ obtained from the Pantheon+SH0ES dataset. The shaded region denotes the $1\sigma$ confidence interval derived from the posterior distribution.

\subsection{Union3 Dataset}

To provide an independent verification of the proposed cosmological models, we further employ the recently released Union3 (UNITY1.5) Type Ia supernova compilation. The Union3 dataset consists of 2087 standardized Type Ia supernovae collected from 24 independent surveys and analyzed using the hierarchical Bayesian UNITY1.5 framework. This compilation provides one of the most comprehensive supernova datasets currently available for precision cosmology.\\

For computational efficiency, the Union3 collaboration provides a compressed likelihood consisting of 22 binned distance modulus measurements together with the corresponding covariance matrix. The compressed dataset spans the redshift interval

\begin{equation}
0.05 \lesssim z \lesssim 2.26.
\end{equation}

The theoretical luminosity distance is evaluated assuming a spatially flat Friedmann--Lemaître--Robertson--Walker universe,

\begin{equation}
d_L(z)
=
(1+z)c
\int_0^z
\frac{dz'}{H(z')},
\end{equation}

where $c$ is the speed of light and $H(z)$ denotes the Hubble parameter predicted by the cosmological model.

The corresponding theoretical distance modulus is given by

\begin{equation}
\mu_{\rm th}(z)
=
5\log_{10}
\left(
\frac{d_L(z)}{\rm Mpc}
\right)
+25.
\end{equation}

\begin{figure}[H]
    \centering
      \includegraphics[width=0.8199\linewidth]
      {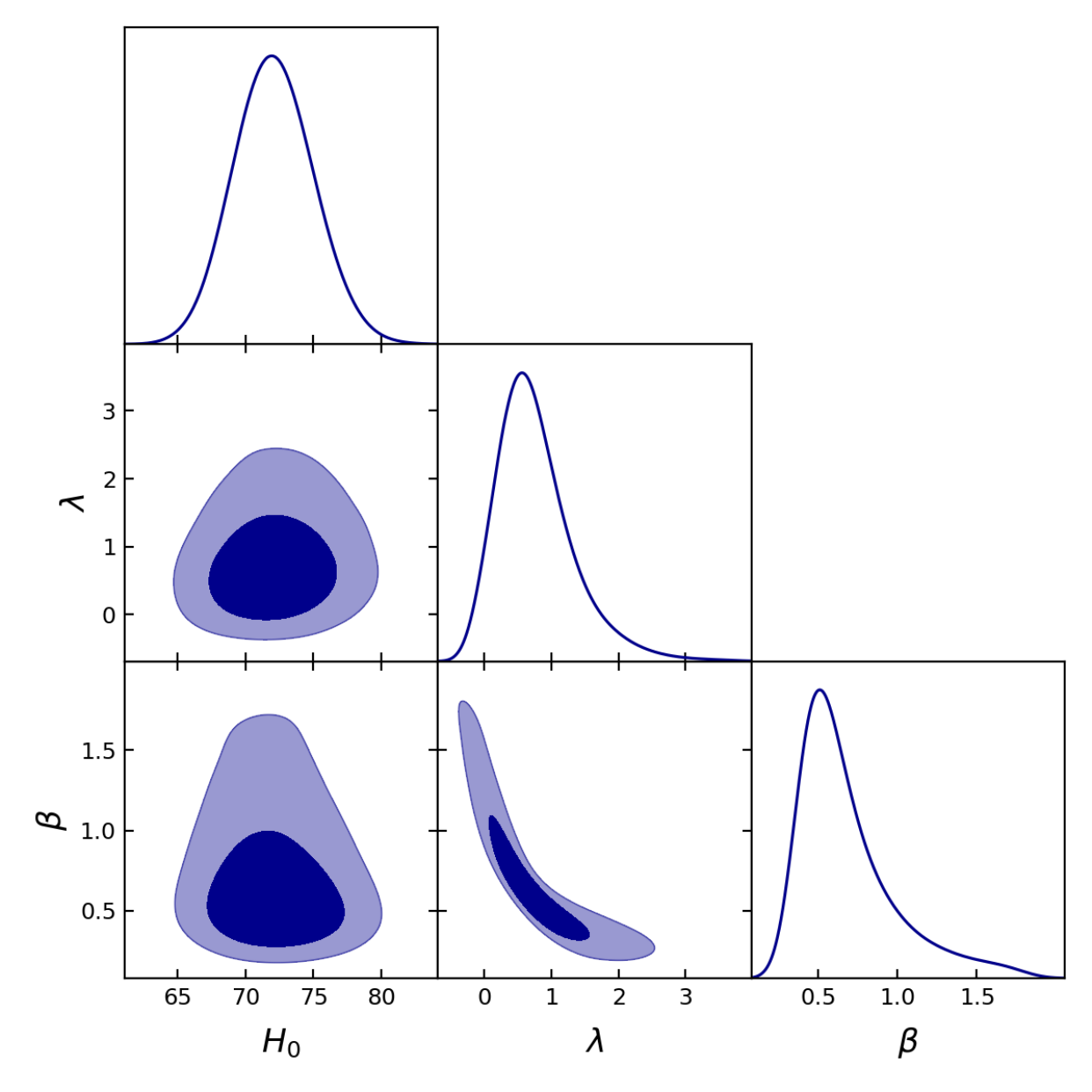}
    \caption{Marginalized posterior distributions and posterior contours for the free parameters $(H_{0},\,\lambda,\,\beta)$ obtained from the Union3 Type Ia supernova dataset. The diagonal panels show the one-dimensional marginalized posterior distributions, whereas the off-diagonal panels present the corresponding two-dimensional posterior contours. The shaded regions denote the 68\% and 95\% Bayesian credible regions inferred from the MCMC analysis.
}
    \label{fig: Union3}
\end{figure}

The agreement between the theoretical predictions and the Union3 observations is
quantified using the full covariance likelihood,

\begin{equation}
\chi^2_{\rm Union3}
=
\Delta\boldsymbol{\mu}^{\,T}
\mathbf{C}^{-1}
\Delta\boldsymbol{\mu},
\end{equation}

where

\begin{equation}
\Delta\boldsymbol{\mu}
=
\mu_{\rm th}
-
\mu_{\rm obs},
\end{equation}

and $\mathbf{C}^{-1}$ denotes the inverse covariance matrix associated with the
compressed Union3 distance modulus measurements. We adopt the official UNITY1.5
compressed release of Rubin et al.~\cite{Rubin:2023jdq}, consisting of 22 binned distance moduli together with their $22\times22$
inverse covariance matrix (combined statistical and systematic uncertainties), The inverse covariance matrix is used directly in the likelihood evaluation of Eq.~(50), without requiring an additional matrix inversion.Fig.~\ref{fig: Union3}\\

\subsection{Combined Analysis}

To obtain the final constraints on the proposed hybrid cosmological model, we perform a joint likelihood analysis by combining the Cosmic Chronometer (CC), DESI DR2 BAO, Pantheon+SH0ES, and Union3 datasets. Assuming that these observational probes are statistically independent, the total likelihood is constructed as the product of the individual likelihood functions, which is equivalent to minimizing the sum of their corresponding chi-square functions,

\begin{equation}
\chi_{\rm Combined}^{2}
=
\chi_{\rm CC}^{2}
+
\chi_{\rm BAO}^{2}
+
\chi_{\rm Pantheon+SH0ES}^{2}
+
\chi_{\rm Union3}^{2}.
\end{equation}

\begin{figure}[H]
    \centering
      \includegraphics[width=0.8199\linewidth]
      {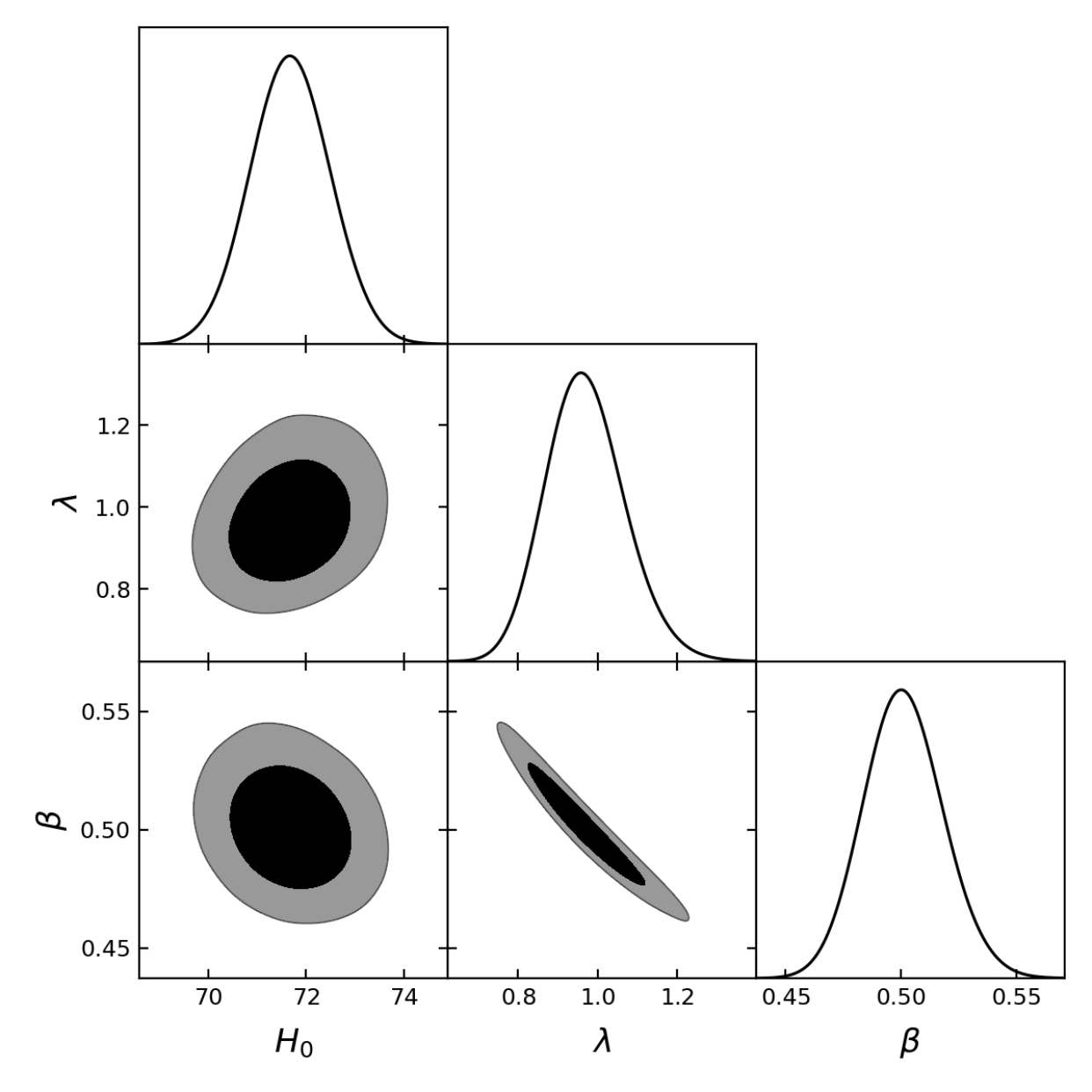}
   \caption{Marginalized posterior distributions and posterior contours for the free parameters $(H_{0},\,\lambda,\,\beta)$ obtained from the combined analysis of the Cosmic Chronometer (CC), DESI DR2 BAO, Pantheon+SH0ES, and Union3 datasets. The diagonal panels show the one-dimensional marginalized posterior distributions, whereas the off-diagonal panels present the corresponding two-dimensional posterior contours. The shaded regions denote the 68\% and 95\% Bayesian credible regions inferred from the MCMC analysis.
}
    \label{fig: Combined}
\end{figure}

\begin{figure}[H]
    \centering
      \includegraphics[width=0.9999\linewidth]
      {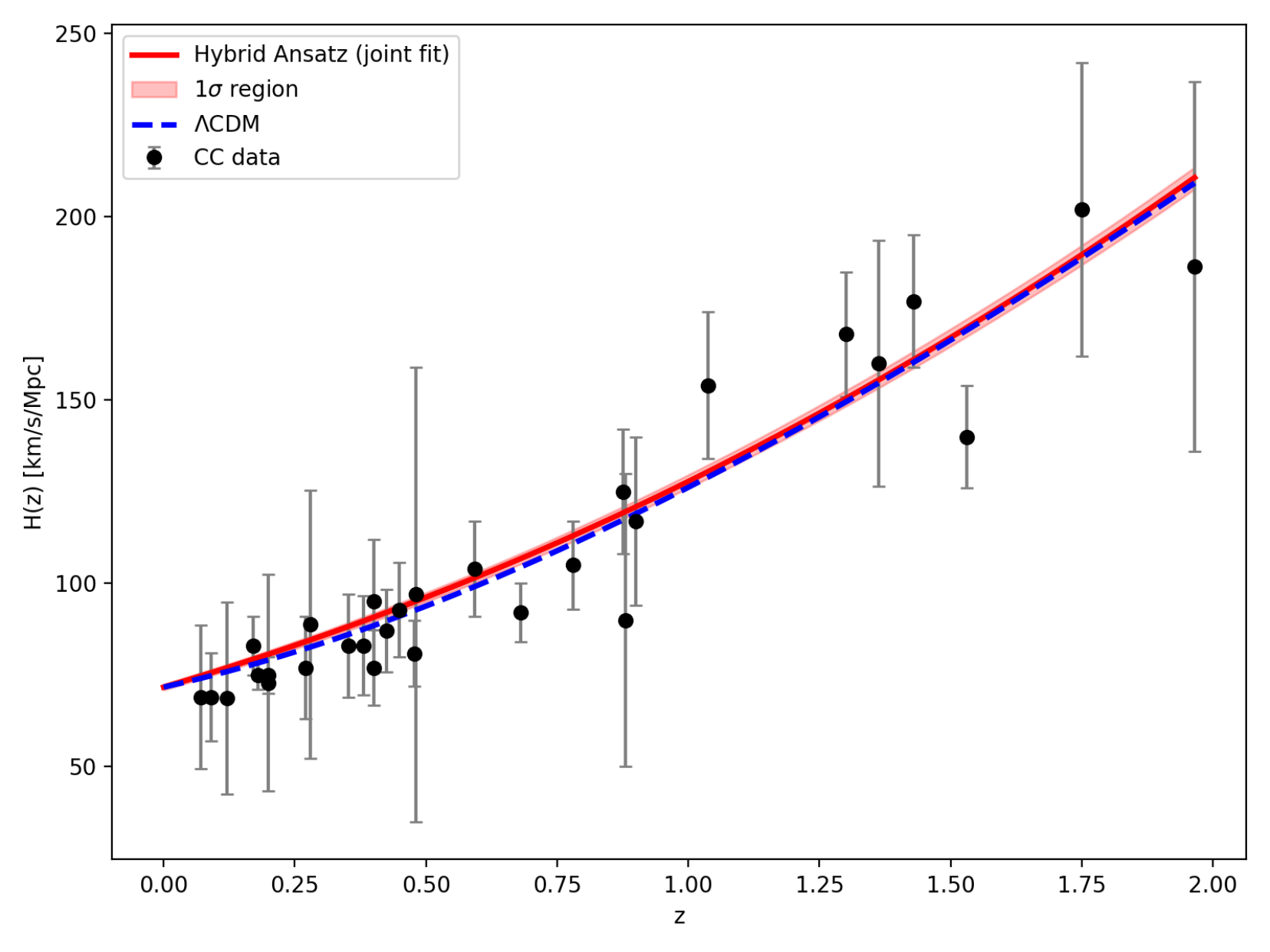}
    \caption{Best-fit hybrid ansatz model compared with the Cosmic Chronometer (CC) data from the joint analysis. The red shaded region denotes the $1\sigma$ confidence interval, while the dashed blue curve represents the $\Lambda$CDM model.}
    \label{fig: Joint}
\end{figure}

The corresponding marginalized posterior distributions and confidence contours obtained from the combined analysis are presented in Fig.~\ref{fig: Combined} and \ref{fig: Joint}, while the resulting best-fit parameter values are summarized in Table~\ref{tab:mcmc_results}.

\begin{table*}[ht]
\centering
\caption{Best-fit values of the model parameters obtained from the MCMC analysis of different observational datasets. The uncertainties correspond to the $68\%$ confidence interval\\.}
\label{tab:mcmc_results}
\begin{tabular}{lccc}
\toprule
Dataset & $H_0$ (km s$^{-1}$ Mpc$^{-1}$) & $\lambda$ & $\beta$ \\
\midrule
CC &
$66.1576^{+3.6733}_{-3.9121}$ &
$0.5252^{+0.7012}_{-0.3957}$ &
$0.6318^{+0.2025}_{-0.1438}$ \\
\\
Union3 &
$72.0363^{+2.9942}_{-2.8870}$ &
$0.6691^{+0.6182}_{-0.4501}$ &
$0.5899^{+0.3651}_{-0.1842}$ \\
\\
BAO &
$67.1853^{+0.7489}_{-0.7665}$ &
$0.8511^{+0.1510}_{-0.1345}$ &
$0.5203^{+0.0283}_{-0.0262}$ \\
\\
Pantheon+SH0ES &
$73.2874^{+1.0226}_{-0.9894}$ &
$1.1438^{+0.5813}_{-0.4405}$ &
$0.4637^{+0.1672}_{-0.1019}$ \\
\\
Combined &
$\mathbf{71.6773}^{+0.9655}_{-0.5009}$ &
$\mathbf{0.9655}^{+0.1023}_{-0.0933}$ &
$\mathbf{0.5009}^{+0.0174}_{-0.0167}$ \\
\bottomrule
\end{tabular}

\end{table*}

\section{Physical and dynamical Study}

\subsection{Deceleration Parameter}

The deceleration parameter is one of the most fundamental cosmological quantities used to characterize the expansion history of the universe. It provides a direct measure of whether the cosmic expansion is accelerating or decelerating and is defined as

\begin{equation}
q(z)=-1+\frac{(1+z)}{H(z)}\frac{dH(z)}{dz}.
\label{eq:q_definition}
\end{equation}

For the proposed hybrid Lambert-W reconstruction, the above expression admits the following analytical form

\begin{equation}
q(z)=\frac{1}{\beta\left[1+W(x)\right]^2}-1,
\label{eq:q_hybrid}
\end{equation}

where

\begin{equation}
x=\frac{\lambda}{\beta}(1+z)^{-1/\beta},
\end{equation}

and $W(x)$ denotes the principal branch of the Lambert-W function. An important feature of Eq.~(\ref{eq:q_hybrid}) is that the Hubble constant $H_0$ does not explicitly appear in the expression, indicating that the deceleration parameter depends solely on the shape of the expansion history governed by the model parameters $\lambda$ and $\beta$.\\

\begin{figure}[H]
   \centering
     \includegraphics[width=0.9999\linewidth]{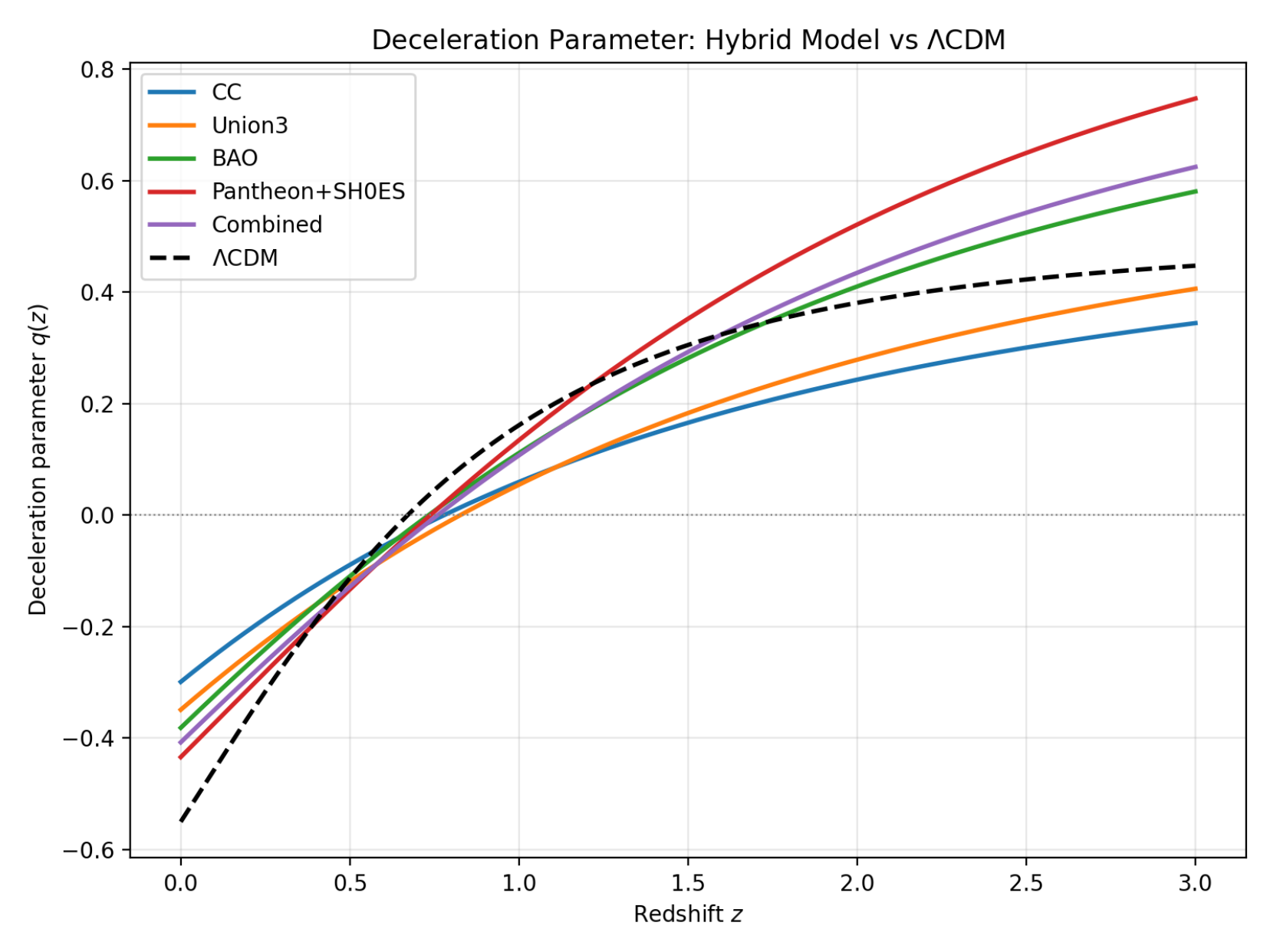}
    \caption{Evolution of the deceleration parameter $q(z)$ for the proposed hybrid cosmological model using the best-fit parameters obtained from the CC, Union3, BAO, Pantheon+SH0ES, and combined datasets. The dashed black curve represents the $\Lambda$CDM prediction.}
    \label{fig: deceleration}
\end{figure}

Figure~\ref{fig: deceleration} illustrates the evolution of the deceleration parameter for the best-fit values obtained from the individual observational datasets, namely CC, Union3, BAO, Pantheon+SH0ES, and the combined analysis. For comparison, the prediction of the standard $\Lambda$CDM model with $\Omega_{m0}=0.30$ and $H_0=68~\mathrm{km\,s^{-1}\,Mpc^{-1}}$ is also shown.\\

At the present epoch ($z=0$), all reconstructed models yield negative values of the deceleration parameter, confirming that the Universe is currently undergoing accelerated expansion. Specifically, the present-day deceleration parameters are found to be $q_0=-0.2992$ for the CC dataset, $q_0=-0.3492$ for Union3, $q_0=-0.3815$ for BAO, $q_0=-0.4340$ for Pantheon+SH0ES, and $q_0=-0.4076$ for the combined dataset. The corresponding transition redshifts, marking the epoch at which the Universe evolves from decelerated to accelerated expansion, are $z_t=0.7788$, $0.8266$, $0.7330$, $0.7384$, and $0.7592$, respectively. For comparison, the standard $\Lambda$CDM model predicts $q_0=-0.5500$ with a transition redshift of $z_t=0.6711$. Although the reconstructed models exhibit slightly less negative present-day deceleration parameters than the $\Lambda$CDM prediction, they consistently reproduce the observed late-time accelerated expansion and yield transition redshifts that are broadly consistent with the standard cosmological scenario.\\

A comparison among the different datasets reveals that the overall behaviour of the reconstructed models remains qualitatively similar despite slight quantitative differences in the best-fit parameters. The combined observational analysis closely follows the behaviour predicted by the individual datasets while remaining reasonably consistent with the $\Lambda$CDM model over the entire redshift interval. Small deviations between the reconstructed model and $\Lambda$CDM become more noticeable at higher redshifts; however, these differences arise from the modified expansion dynamics introduced through the hybrid Lambert-W parameterization rather than indicating any inconsistency with the observational constraints. Overall, the evolution of the deceleration parameter demonstrates that the proposed hybrid reconstruction successfully captures the key kinematic features of the observed cosmic expansion history. In particular, the model reproduces the transition from a decelerating matter-dominated universe to the present phase of accelerated expansion while maintaining close agreement with the concordance $\Lambda$CDM cosmology.

\subsection{Equation of state parameter}

The effective equation of state (EoS) parameter provides valuable insight into the dynamical evolution of the universe by characterizing the overall pressure-to-energy density ratio of the cosmic fluid. Unlike the equation of state associated with an individual matter component, the effective EoS incorporates the combined influence of all contributions driving the cosmic expansion. Consequently, it serves as a convenient diagnostic for distinguishing between accelerating and decelerating cosmological phases in modified gravity theories, including $f(T)$ gravity~\cite{Cai2016}. The effective equation-of-state parameter is defined as,

\begin{equation}
w_{\rm eff}(z)=\frac{p}{\rho},
\label{eq:weff}
\end{equation}

\begin{figure}[H]
   \centering
       \includegraphics[width=0.8199\linewidth]{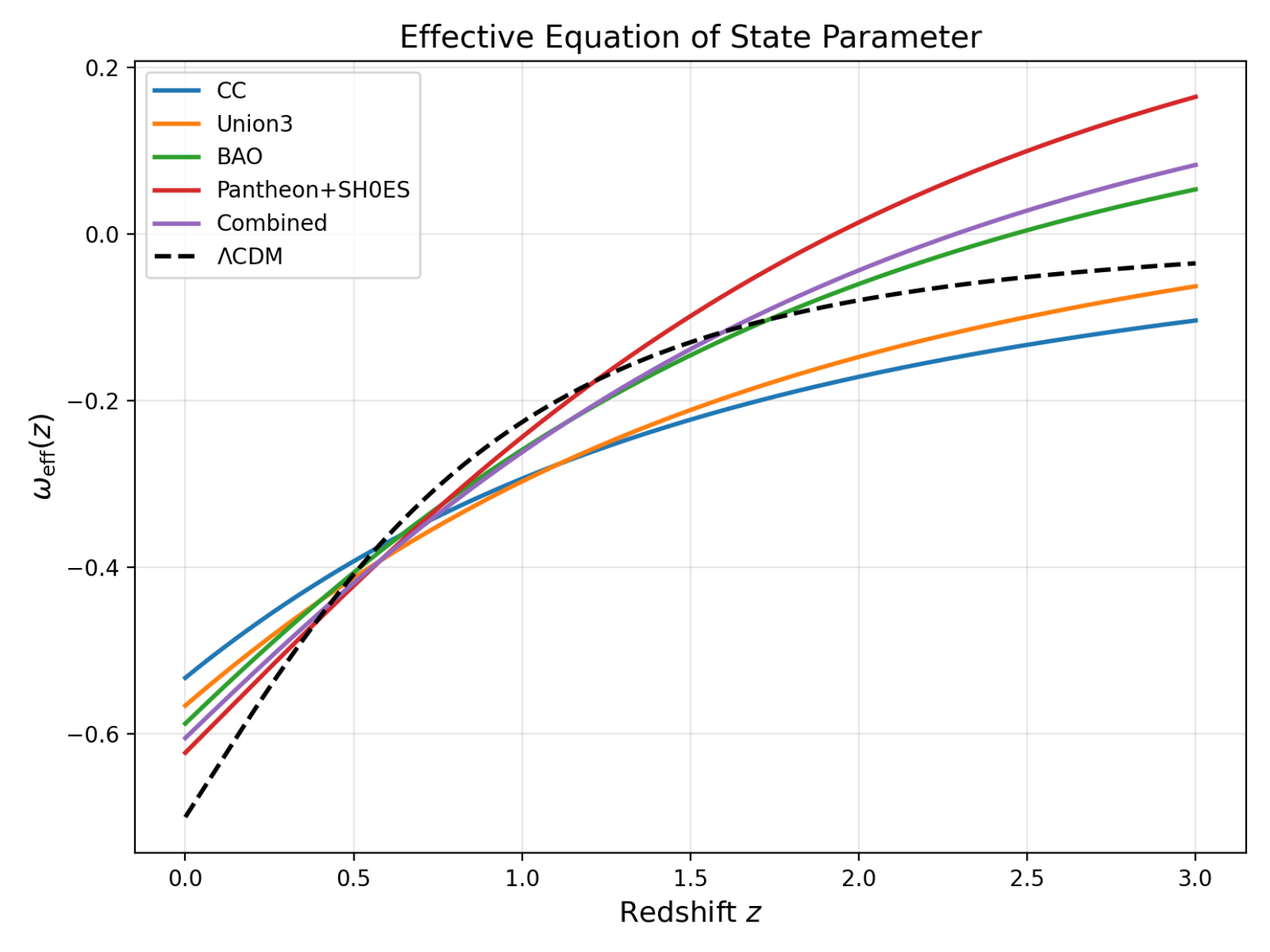}
       \caption{Redshift evolution of the effective equation-of-state parameter $w_{\rm eff}(z)$ for the proposed hybrid cosmological model using the best-fit parameters constrained by the CC, Union3, BAO, Pantheon+SH0ES, and combined datasets. The dashed black curve corresponds to the $\Lambda$CDM model.}
    \label{fig:Equation of state}
\end{figure}

Figure \ref{fig:Equation of state} compares the redshift evolution of the effective equation of state corresponding to the best-fit parameters obtained from the individual observational datasets, namely CC, Union3, BAO, Pantheon+SH0ES, and the combined analysis. For comparison, the prediction of the standard $\Lambda$CDM cosmology is also shown. At the present epoch ($z=0$), the effective equation of state remains negative for all reconstructed models, confirming that the proposed hybrid cosmological model successfully reproduces the observed late-time accelerated expansion of the Universe. Specifically, the present-day values are found to be $w_{\rm eff}(0)=-0.5328$ for the CC dataset, $-0.5661$ for Union3, $-0.5877$ for BAO, $-0.6227$ for Pantheon+SH0ES, and $-0.6051$ for the combined dataset. For comparison, the standard $\Lambda$CDM model predicts $w_{\rm eff}(0)=-0.7000$. Although the reconstructed models exhibit slightly higher values of the effective equation of state than the $\Lambda$CDM prediction, they consistently remain within the quintessence regime ($-1<w_{\rm eff}<-1/3$), indicating that the cosmic acceleration is driven by an effective dark-energy component without crossing the phantom divide. Furthermore, the overall evolution of $w_{\rm eff}(z)$ remains qualitatively consistent across all observational datasets, demonstrating the robustness of the proposed hybrid reconstruction in describing the late-time expansion history of the Universe.\\

A comparison among the various observational datasets reveals that the overall behaviour of the effective equation of state remains remarkably consistent despite the slight differences in the best-fit values of the model parameters. The combined dataset closely follows the trends exhibited by the individual analyses and remains in reasonable agreement with the $\Lambda$CDM prediction over the considered redshift range. Minor deviations observed at intermediate and higher redshifts originate from the modified expansion dynamics introduced through the hybrid Lambert-W parameterization and reflect the additional flexibility of the reconstructed model rather than any incompatibility with observational data. Overall, the evolution of the effective equation of state demonstrates that the proposed hybrid reconstruction successfully reproduces the essential features of the observed cosmic expansion history. The model predicts a present-day accelerated universe characterized by a negative effective equation of state, remains within the physically viable quintessence regime, and exhibits a smooth transition toward the matter-dominated epoch at earlier times.

\subsection{Statefinder parameter}

The statefinder diagnostic pair $(r,s)$ provides a powerful geometrical tool for examining the validity of cosmological models. It extends the analysis beyond basic kinematic quantities by incorporating higher-order derivatives of the scale factor. In particular, the parameter $r$, known as the jerk, involves the third derivative of the scale factor, while $s$ is constructed using both $r$ and the deceleration parameter $q$. Since these parameters depend only on the scale factor and its derivatives, they are purely kinematic in nature and depend exclusively on the metric structure, making them independent of any specific gravitational theory.

The statefinder parameters can be written in terms of the Hubble parameter as follows\cite{Sahni2003},
\begin{equation}
    r=\frac{\dddot a}{a H^3}, \qquad
    s=\frac{r-1}{3\left(q-\frac{1}{2}\right)}
\end{equation}

The statefinder pair $(r,s)$ serves as an effective diagnostic to characterize different phases of cosmic evolution. The fixed point $(r,s)=(1,0)$ corresponds to the standard spatially flat $\Lambda CDM$ model, which acts as a reference point in cosmology. Models with $r<1$ and $s>0$ represent a quintessence-like phase, whereas the point $(r,s)=(1,1)$ is also associated with the standard cold dark matter (SCDM) model under suitable conditions.\newline

\begin{figure}[H]
  \centering
    \includegraphics[width=0.8199\linewidth]{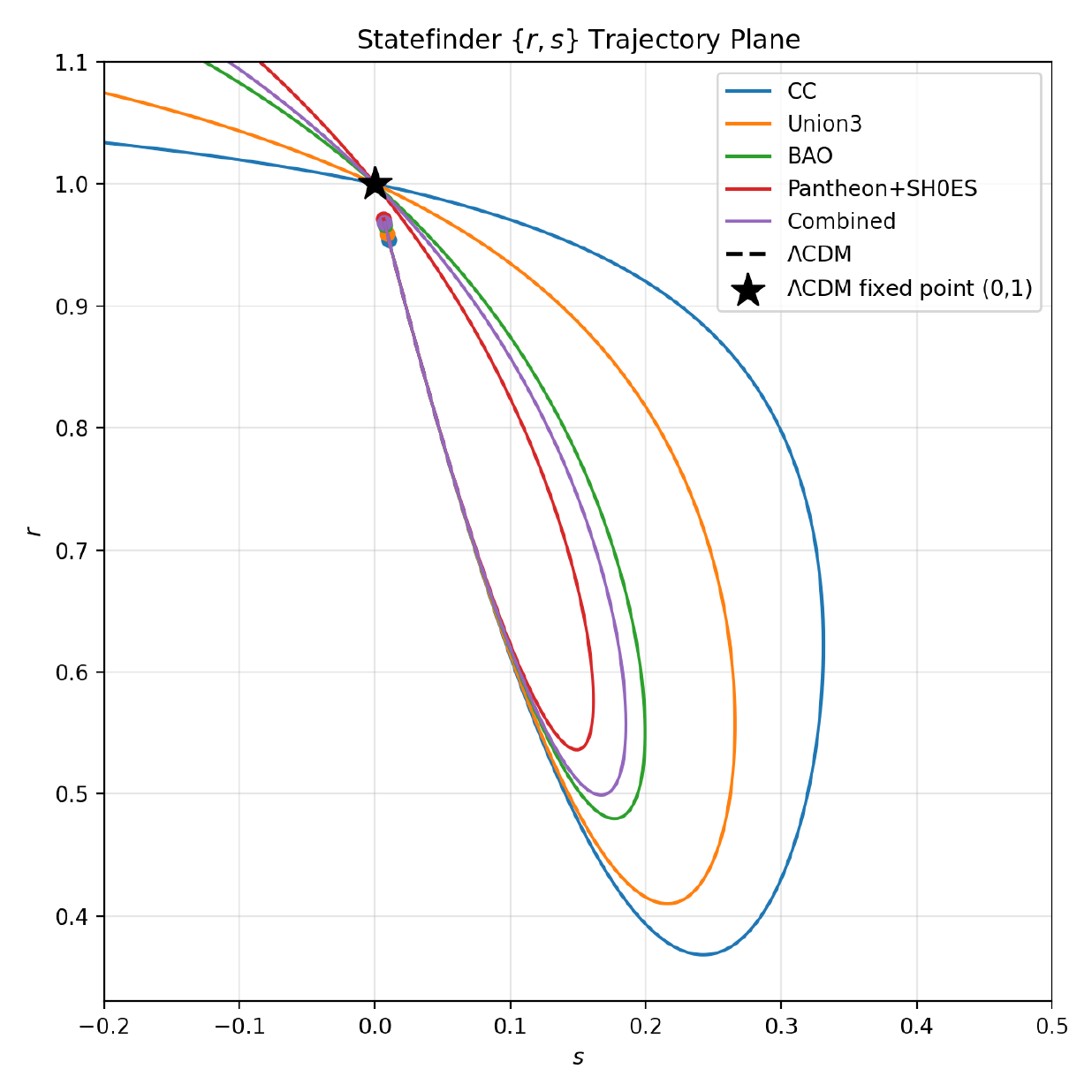}
   \caption{Statefinder trajectories in the $\{r,s\}$ plane for the proposed hybrid cosmological model corresponding to the best-fit parameters obtained from the CC, Union3, BAO, Pantheon+SH0ES, and combined datasets. The black star at $(s,r)=(0,1)$ denotes the fixed point of the $\Lambda$CDM model.}
   \label{statefinder}
  \end{figure}

 Figure \ref{statefinder} presents the evolution of the statefinder trajectories obtained using the CC, Union3, BAO, Pantheon+SH0ES, and combined observational datasets. It is evident that all trajectories originate in the quintessence region, indicating that the proposed hybrid model initially exhibits a dynamical dark energy behavior rather than a pure cosmological constant. The Pantheon+SH0ES dataset shows the largest departure from the $\Lambda$CDM fixed point during the early stage of evolution, while the CC, Union3, BAO, and combined datasets display comparatively smaller deviations. As the cosmic evolution proceeds, the trajectories gradually converge toward the fixed point $(r,s)=(1,0)$, demonstrating that the model asymptotically approaches the standard $\Lambda$CDM scenario at late times. Although minor differences are observed among the individual datasets, the overall evolution remains qualitatively similar, indicating that the hybrid model provides a stable and observationally consistent description of the accelerated expansion of the universe. The close agreement of the combined dataset with the individual observational constraints further supports the robustness of the proposed cosmological model.

\subsection{Om(z) Diagnostic}

Unlike cosmological diagnostics that involve higher-order derivatives of the scale factor, the $Om(z)$ diagnostic depends solely on the Hubble expansion rate and therefore provides a relatively model-independent test of the underlying cosmological dynamics. One of its most important features is that it remains constant for the spatially flat $\Lambda$CDM model, whereas any deviation from this constant behaviour indicates departures from the standard cosmological scenario.\\

The $Om(z)$ diagnostic is defined as\cite{Sahni:2008xx}

\begin{equation}
Om(z)=
\frac{E^{2}(z)-1}
{(1+z)^{3}-1},
\label{eq:Om}
\end{equation}

where

\begin{equation}
E(z)=\frac{H(z)}{H_{0}}
\end{equation}

is the normalized Hubble parameter. For the spatially flat $\Lambda$CDM model, Eq.~(\ref{eq:Om}) reduces to

\begin{equation}
Om(z)=\Omega_{m0},
\end{equation}

which is independent of redshift. Consequently, any redshift evolution of $Om(z)$ directly reflects deviations from the concordance cosmological model.\\

\begin{figure}[H]
    \centering
      \includegraphics[width=0.9999\linewidth]{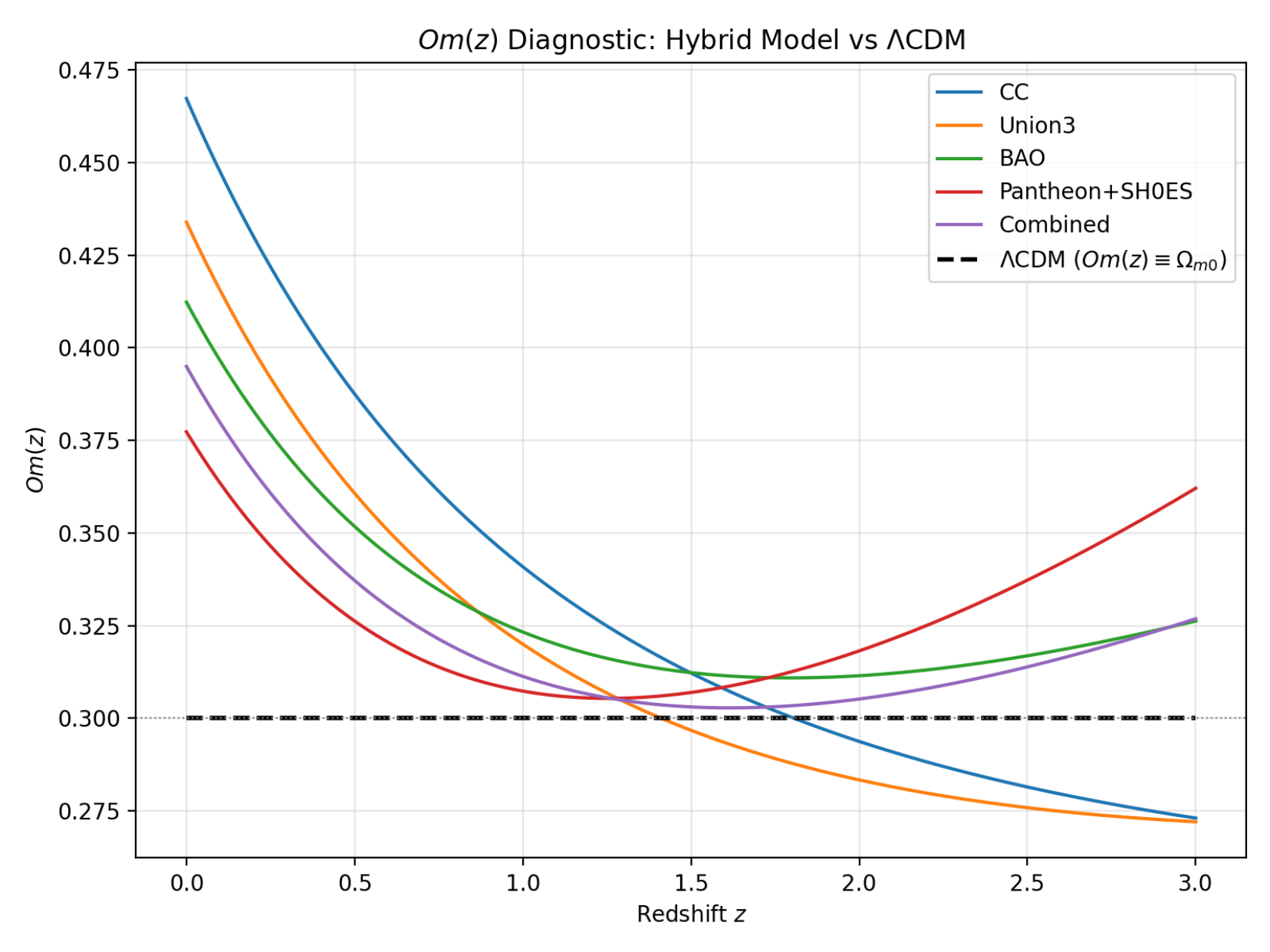}
    \caption{Evolution of the $Om(z)$ diagnostic for the proposed hybrid model using the best-fit parameters obtained from different observational datasets. The horizontal dashed line denotes the constant $\Lambda$CDM prediction, $Om(z)=\Omega_{m0}=0.30$.}
  \label{fig:omz}  
\end{figure}

Figure~\ref{fig:omz} presents the evolution of the $Om(z)$ diagnostic for the best-fit parameters obtained from the CC, Union3, BAO, Pantheon+SH0ES, and combined observational datasets. For comparison, the corresponding prediction of the standard $\Lambda$CDM cosmology with $\Omega_{m0}=0.30$ is also shown. As expected, the $\Lambda$CDM model produces a constant horizontal line throughout the entire redshift interval, confirming the well-known property that the $Om(z)$ diagnostic is independent of redshift in the concordance model.\\

The reconstructed hybrid model exhibits a smooth and continuous evolution of the $Om(z)$ diagnostic for all observational datasets, with no evidence of oscillatory or singular behaviour throughout the investigated redshift interval. Although the overall evolution differs among the individual best-fit solutions owing to the variations in the optimized parameters $(H_0,\lambda,\beta)$, all reconstructed curves remain finite and physically well behaved. The CC and Union3 datasets display a monotonic decrease in $Om(z)$ with increasing redshift, approaching and eventually falling below the $\Lambda$CDM value at higher redshifts. In contrast, the BAO, Pantheon+SH0ES, and combined datasets exhibit a shallow minimum around intermediate redshifts followed by a gradual increase towards higher redshifts. Such behaviour reflects the modified expansion dynamics introduced by the hybrid Lambert--W reconstruction while preserving a stable cosmological evolution.\\

A quantitative comparison with the $\Lambda$CDM prediction shows that the reconstructed models generally possess larger $Om(z)$ values than the constant reference value $\Omega_{m0}=0.30$ at low redshifts. For example, at $z=0$ the reconstructed values range from $Om(0)=0.3773$ (Pantheon+SH0ES) to $Om(0)=0.4672$ (CC), whereas the $\Lambda$CDM model remains fixed at $Om(z)=0.30$. As the redshift increases, the deviations become progressively smaller, with several reconstructed curves approaching the $\Lambda$CDM prediction around $z\approx1.5$--$2.0$. Beyond this epoch, the CC and Union3 reconstructions decrease slightly below the $\Lambda$CDM value, while the BAO, Pantheon+SH0ES, and combined datasets remain marginally above it. These smooth redshift-dependent deviations are characteristic of the modified expansion dynamics of the reconstructed exponential $f(T)$ model and do not indicate any pathological behaviour. \\

\subsection{Energy Conditions}

The energy conditions provide an important theoretical consistency test for any cosmological model by imposing physically motivated constraints on the effective energy density and pressure governing the cosmic expansion. These conditions are widely employed in both general relativity and modified gravity theories to examine whether the effective cosmic fluid satisfies the fundamental requirements associated with attractive gravity, causal propagation of energy, and positive energy density\cite{Kontou:2020bta}. Within the framework of the reconstructed hybrid model, the effective energy density and pressure are obtained from the modified Friedmann equations and are expressed in terms of the normalized Hubble parameter and the effective equation of state as

\begin{equation}
\rho_{\rm eff}=3E^{2}(z),
\label{eq:rhoeff}
\end{equation}

and

\begin{equation}
p_{\rm eff}=w_{\rm eff}(z)\rho_{\rm eff},
\label{eq:peff}
\end{equation}\\

where $E(z)=H(z)/H_0$ denotes the normalized Hubble parameter. Using these quantities, the standard energy conditions can be written as

\begin{align}
{\rm NEC}: \qquad
&\rho_{\rm eff}+p_{\rm eff}\ge0,\\
{\rm WEC}: \qquad
&\rho_{\rm eff}\ge0,\qquad
\rho_{\rm eff}+p_{\rm eff}\ge0,\\
{\rm SEC}: \qquad
&\rho_{\rm eff}+3p_{\rm eff}\ge0,\qquad
\rho_{\rm eff}+p_{\rm eff}\ge0,\\
{\rm DEC}: \qquad
&\rho_{\rm eff}\ge0,\qquad
\rho_{\rm eff}\pm p_{\rm eff}\ge0.
\end{align}

\begin{figure}[H]
   \centering
     \includegraphics[width=0.8999\linewidth]{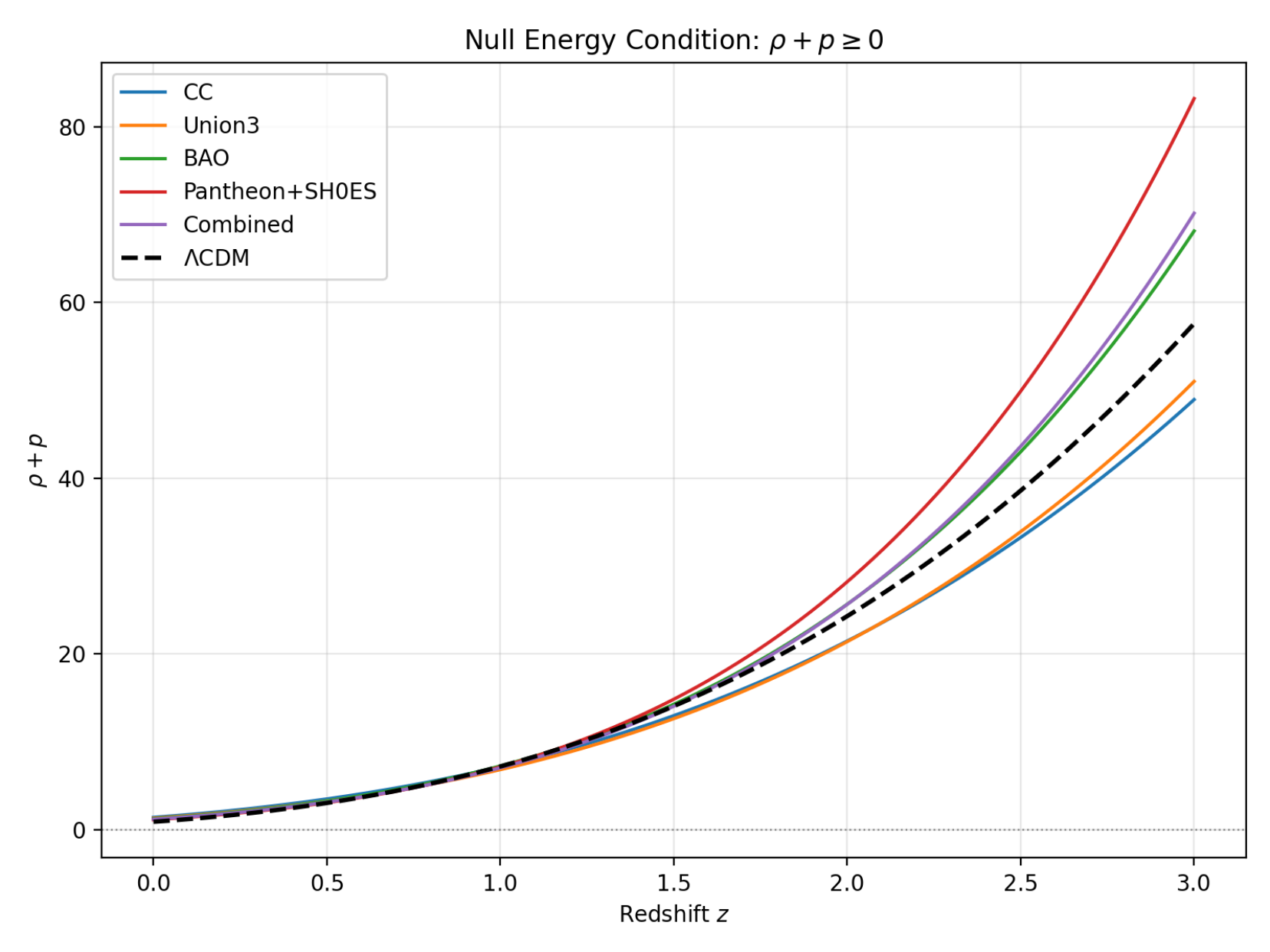}
   \caption{Evolution of the Null Energy Condition (NEC), $\rho+p\geq0$, for the proposed hybrid cosmological model using the best-fit parameters constrained by different observational datasets.}
   \label{fig:NEC}
   \end{figure}

\begin{figure}[H]
   \centering
    \includegraphics[width=0.8999\linewidth]{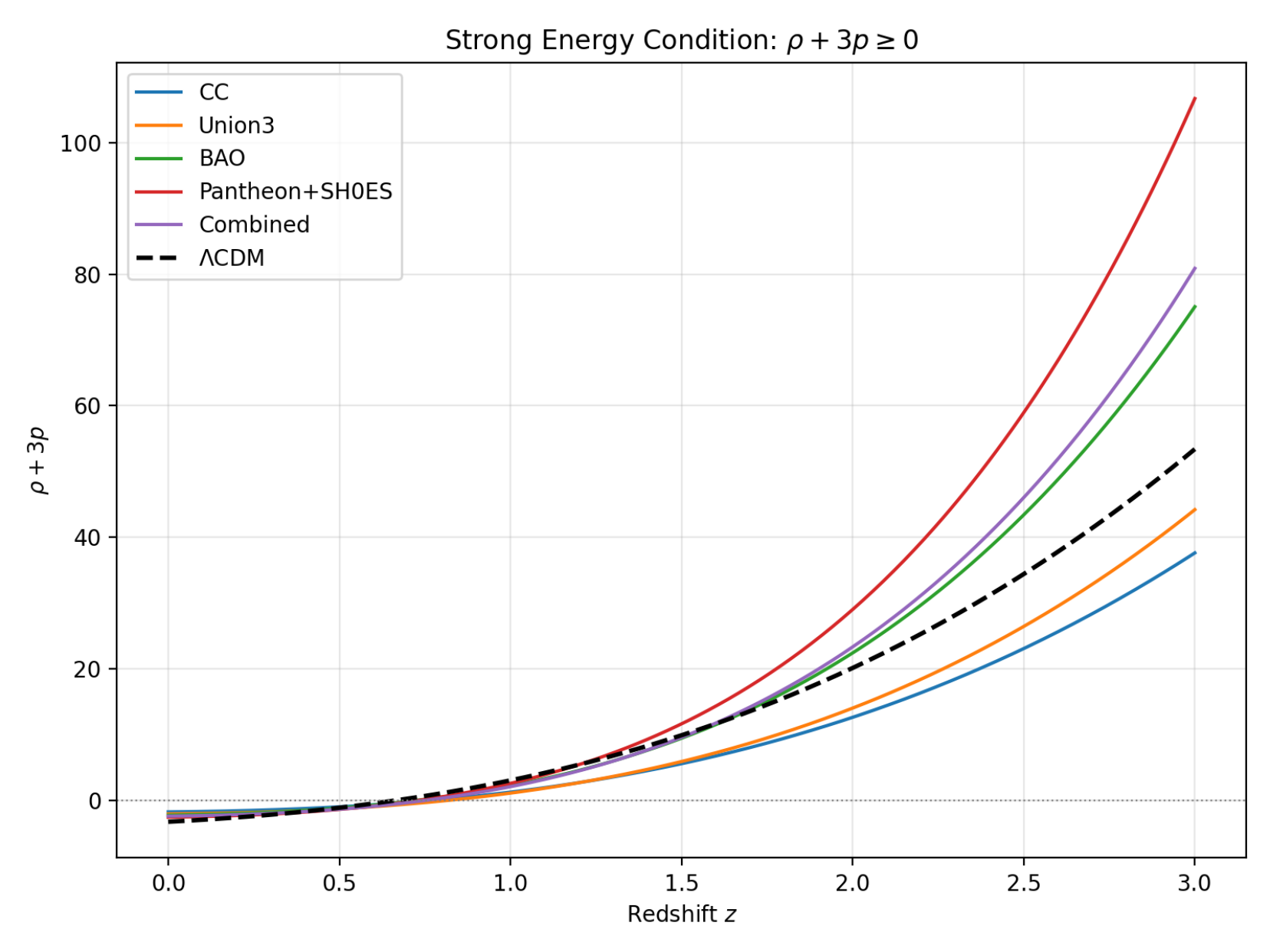}
   \caption{Evolution of the Strong Energy Condition (SEC), $\rho+3p\geq0$, for the proposed hybrid cosmological model using the best-fit parameters constrained by different observational datasets.}
   \label{fig:SEC}
   \end{figure}

\begin{figure}[H]
   \centering
     \includegraphics[width=0.8999\linewidth]{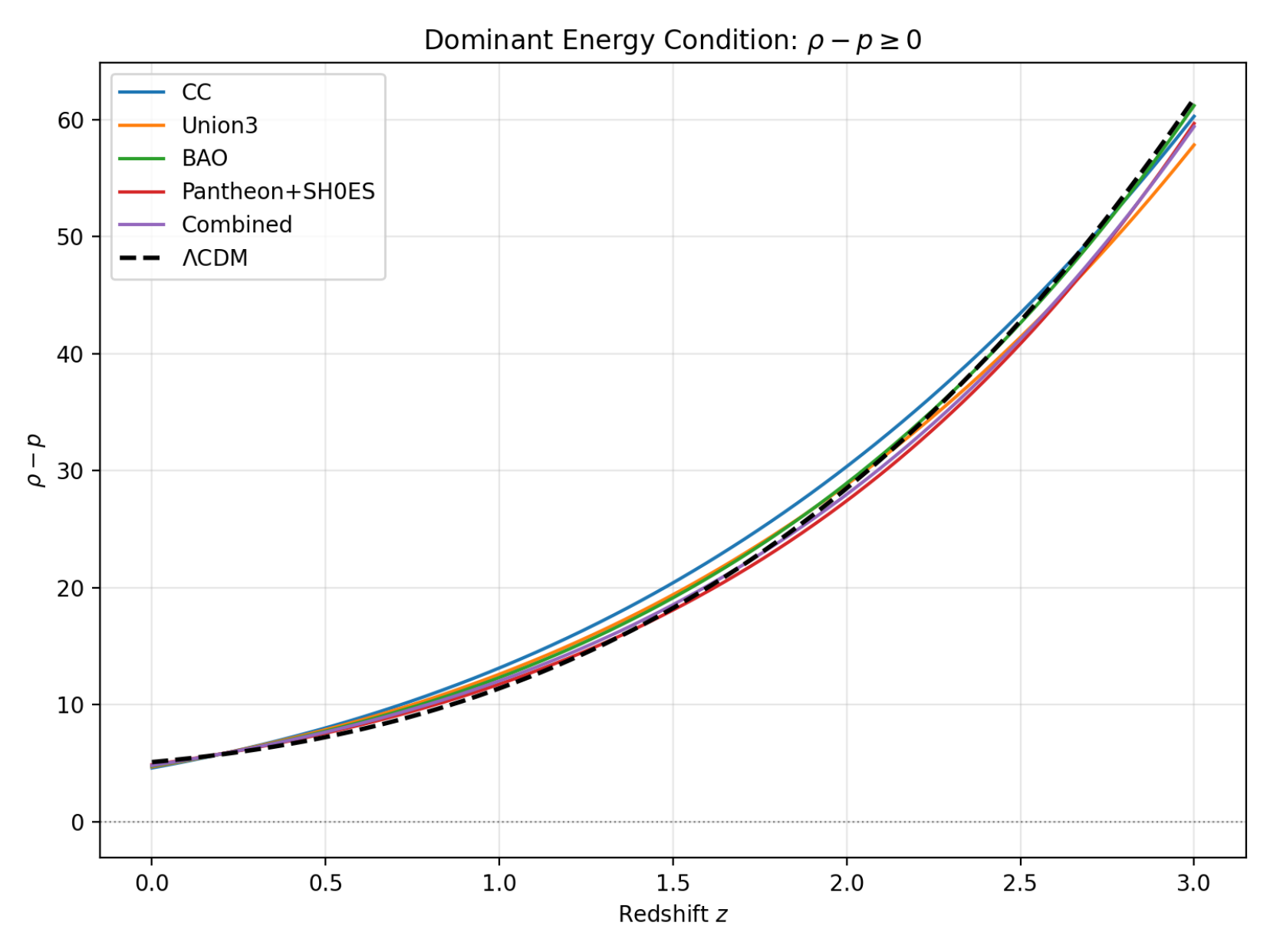}
   \caption{Evolution of the Dominant Energy Condition (DEC), $\rho-p\geq0$, for the proposed hybrid cosmological model using the best-fit parameters constrained by different observational datasets.}
   \label{fig:DEC}
   \end{figure}

The evolution of these energy conditions corresponding to the best-fit parameters obtained from the Cosmic Chronometer (CC), Union3, BAO, Pantheon+SH0ES, and combined observational datasets is illustrated in Figs.~\ref{fig:NEC}, \ref{fig:SEC}, and \ref{fig:DEC}, respectively. For comparison, the corresponding predictions of the standard $\Lambda$CDM cosmology are also shown.\\

Figure~\ref{fig:NEC} presents the behaviour of the Null Energy Condition (NEC). It is evident that the quantity $\rho_{\rm eff}+p_{\rm eff}$ remains positive throughout the entire redshift interval for all observational datasets. The absence of any violation indicates that the reconstructed hybrid model satisfies the null energy condition over the entire cosmological evolution considered in this work. Furthermore, the reconstructed curves closely follow the behaviour of the $\Lambda$CDM model, with the reconstructed curves exhibiting a similar overall evolution to the $\Lambda$CDM prediction while displaying gradually increasing deviations towards higher redshifts as a consequence of the modified expansion dynamics introduced by the hybrid Lambert--W parametrization. The Weak Energy Condition (WEC) is also satisfied throughout the investigated redshift interval. Since the effective energy density remains strictly positive, $\rho_{\rm eff}=3E^{2}(z)>0$, for all redshifts considered, and the Null Energy Condition is simultaneously fulfilled, the WEC is automatically satisfied for all observational datasets. This result confirms that the effective energy density remains positive for every observer and further supports the physical consistency of the reconstructed hybrid model. Moreover, the reconstructed solutions remain in good agreement with the corresponding ΛCDM prediction, indicating that the hybrid Lambert--W reconstruction preserves the physical consistency of the effective cosmic fluid throughout the investigated redshift interval.\\

The Strong Energy Condition (SEC), shown in Fig.~\ref{fig:SEC}, exhibits a markedly different behaviour. At the present epoch, the quantity $\rho_{\rm eff}+3p_{\rm eff}$ assumes negative values for all datasets, indicating a clear violation of the SEC. Such a violation is not only expected but is a necessary condition for the observed accelerated expansion of the universe. As the redshift increases, the quantity gradually becomes positive, the effective cosmic fluid once again satisfies the strong energy condition, signalling the transition to the decelerating matter-dominated phase. The transition from negative to positive values therefore naturally reflects the change from the present accelerated phase to the decelerating expansion that characterizes the early universe. The transition occurs smoothly for all observational datasets and remains consistent with the behaviour predicted by the concordance $\Lambda$CDM cosmology.\\

The evolution of the Dominant Energy Condition (DEC) is illustrated in Fig.~\ref{fig:DEC}. The quantity $\rho_{\rm eff}-p_{\rm eff}$ remains strictly positive throughout the investigated redshift interval for all observational datasets, demonstrating that the reconstructed cosmological model satisfies the dominant energy condition during the entire cosmic evolution. The positivity of this quantity implies that the effective energy density exceeds the effective pressure, thereby satisfying the standard causal requirement that the effective energy flux remains non-spacelike. Similar to the NEC, the reconstructed curves exhibit overall agreement with the corresponding $\Lambda$CDM prediction while allowing modest deviations arising from the modified gravitational dynamics. Overall, the analysis of the energy conditions demonstrates that the proposed hybrid Lambert--W reconstruction remains physically consistent throughout the investigated cosmological evolution. The simultaneous satisfaction of the Null, Weak, and Dominant Energy Conditions confirms the positivity and causal behaviour of the effective cosmic fluid, whereas the late-time violation of the Strong Energy Condition naturally accounts for the observed accelerated expansion of the Universe. The subsequent recovery of the Strong Energy Condition at higher redshifts indicates a smooth transition toward the conventional matter-dominated epoch. The overall agreement of the reconstructed model with the corresponding $\Lambda$CDM behaviour further supports the physical viability of the proposed analytical reconstruction. The energy conditions corresponding to our model are discussed in the following.\\

\section{Conclusion and Discussion}

In this work, we have presented an analytical reconstruction of the exponential $f(T)$ gravity model by introducing a hybrid expansion ansatz capable of accurately reproducing the late-time cosmological evolution of the original numerical solution. The principal motivation of this reconstruction is to provide an analytical description of the cosmological expansion that eliminates the need for repeated numerical integration of the modified Friedmann equations while preserving the essential physical behaviour of the underlying exponential $f(T)$ model. The reconstructed expansion history was first calibrated to reproduce the numerical solution with high accuracy and subsequently subjected to a comprehensive theoretical and observational investigation to assess its cosmological viability.\\

To constrain the free parameters of the reconstructed model, we employed four independent late-time observational probes, namely Cosmic Chronometers (CC), Baryon Acoustic Oscillations (BAO), Pantheon+SH0ES, and Union3, together with their combined likelihood. The Bayesian parameter estimation was performed using the affine-invariant Markov Chain Monte Carlo (MCMC) technique. The parameter constraints obtained from the individual datasets are mutually consistent, while the combined analysis provides the most stringent bounds owing to the complementary nature of the different cosmological probes. The resulting best-fit parameters provide an excellent description of the observed expansion history over a broad redshift interval, demonstrating the robustness and observational consistency of the proposed analytical reconstruction.\\

The cosmological diagnostics further support the physical consistency of the reconstructed model. The deceleration parameter exhibits the expected transition from an early decelerating phase to the present accelerated epoch, while the effective equation-of-state parameter remains entirely within the quintessence regime throughout the late-time evolution. The $Om(z)$ diagnostic displays mild redshift-dependent deviations from the constant $\Lambda$CDM prediction while remaining in good agreement with the observationally favoured expansion history. Similarly, the statefinder trajectories naturally evolve toward the $\Lambda$CDM fixed point, indicating that the reconstructed model asymptotically approaches the concordance cosmology at late times. Furthermore, the analysis of the energy conditions confirms that the Null, Weak, and Dominant Energy Conditions remain satisfied throughout the investigated cosmic evolution, whereas the late-time violation of the Strong Energy Condition naturally accounts for the observed accelerated expansion of the Universe.\\

The close agreement between the analytical reconstruction, the numerical exponential $f(T)$ model, and the observational constraints demonstrates that the proposed hybrid formulation provides a reliable and computationally efficient alternative to repeated numerical integration of the modified Friedmann equations. The reconstruction preserves the essential characteristics of the original exponential $f(T)$ gravity model while considerably simplifying the cosmological analysis, thereby facilitating parameter estimation and statistical inference using contemporary observational datasets.\\

It is worth emphasizing that the analytical reconstruction presented in this work was initially calibrated by determining the parameters $\lambda$ and $\beta$ such that the reconstructed expansion history accurately reproduces the numerical behaviour of the exponential $f(T)$ model. These parameters were subsequently treated as free parameters and independently constrained through MCMC analyses using the observational datasets considered in this work. The consistency between the reconstructed model, the numerical solution, and the observationally inferred parameter space demonstrates the effectiveness and reliability of the proposed analytical framework.\\

Future investigations may further extend the present framework in several directions. One possible improvement is the development of a more general analytical reconstruction in which the free parameters arise directly from the underlying theoretical formulation, thereby eliminating the need for an initial calibration with the numerical solution. Such an extension could be achieved by introducing additional physically motivated terms or more general functional forms within the hybrid ansatz, allowing the reconstruction parameters to emerge naturally from the theory itself. In addition, the model may be further tested using complementary cosmological probes, including Cosmic Microwave Background (CMB) observations, weak gravitational lensing, redshift-space distortions, baryon growth measurements, and large-scale structure formation data. Such analyses would provide more stringent constraints on the model parameters and enable a more comprehensive assessment of the cosmological viability of the proposed analytical reconstruction.\\

Overall, the present work demonstrates that a suitably constructed analytical reconstruction can faithfully reproduce the cosmological evolution of the exponential $f(T)$ gravity model while remaining consistent with current late-time cosmological observations. Beyond providing a computationally efficient alternative to numerical integration, the proposed hybrid framework offers a practical tool for parameter estimation, cosmological model testing, and theoretical investigations within modified teleparallel gravity. The successful agreement between the analytical reconstruction, the numerical exponential $f(T)$ model, and multiple independent observational datasets establishes the proposed framework as a promising foundation for future analytical studies of modified gravity, dark energy phenomenology, and the accelerated expansion of the Universe.\\

\section*{Acknowledgement}
The authors express sincere thanks to B. Mishra for insightful discussions to improve the manuscript. The authors also sincerely thank the reviewer for the constructive comments and valuable suggestions, which have significantly improved the quality and clarity of the manuscript.

\section*{Declaration}
\textbf{Conflicts of Interest:} The authors have no conflict of interest.
\bibliographystyle{elsarticle-harv}
\balance
\bibliography{reference}

  \end{document}